# The Effect of Water Contamination on the Aging of a Dual-Carbon Lithium-Ion Capacitor Employing LiFSI-Based Electrolyte

Philipp Schweigart[1], Johan Hamonnet[1], Obinna Egwu Eleri[2], Laura King[3], Samson Yuxiu Lai[4], and Ann Mari Svensson[1]

[1]Department of Materials Science and Engineering, Norwegian University of Science and Technology, Trondheim, Norway
[2]Beyonder AS, Sandnes, Norway
[3]Department of Chemistry - Ångström Laboratory, Uppsala University, Uppsala, Sweden
[4]Battery Technology Department, Institute for Energy Technology, Kjeller, Norway

**Corresponding authors**
PS: philipp.schweigart@ntnu.no
AS: annmari.svensson@ntnu.no

# 1 Abstract

Fabricating electrochemical energy storage devices demands significant energy for drying cell components to ensure optimal performance. The development of new, water-tolerant battery materials would represent a tremendous advance in cost savings and sustainability. Although it is generally established that water deteriorates cell performance, there are few systematic studies on the maximum amount of tolerable water contamination, and most of the studies employed the electrolyte salt $LiPF_6$, which inevitably decomposes upon exposure to humidity. In this work, the potential of using the non-hydrolyzing salt LiFSI is explored with respect to its performance in Li-ion capacitor cells based on activated carbon (AC) and pre-lithiated graphite (Gr). Water is deliberately added in various amounts (950, 2300, and 6000 ppm), and its effect on the electrochemical performance and aging of AC and Gr electrodes is systematically studied. Although the addition of 950 ppm water has no evident impact on capacity retention (96% after 2000 cycles), the addition of 2300 ppm water or 6000 ppm water shows a distinct capacity fade, mainly due to a significant loss of lithium inventory and increased resistivity due to irreversible reactions of the water with lithiated Gr. Post-mortem analysis reveals that water promotes the oxidative and reductive decomposition of LiFSI on AC and lithiated Gr, respectively. A significant thickening of the SEI on Gr is observed as the water concentration is increased.

# 2 Introduction

The increase in greenhouse gas emissions and the decades-long reliance of modern society on fossil fuels has accelerated climate change, triggering the rapid development of electrochemical energy storage solutions as alternatives. A wide range of electrochemical energy storage devices that employ different energy storage mechanisms have been implemented to meet the demands of various application fields. Li-ion batteries (LIBs) excel in their attainable energy density by employing faradaic reactions between the host material and $Li^+$ ions primarily via intercalation or alloying reactions. The most widespread anode material in LIBs is graphite (Gr), delivering a capacity of 372 $mAh\ g^{-1}$, whereas transition metal oxides are usually used as cathode material[1]. LIBs are the primary choice for consumer electronics and electric vehicles due to their favorable energy densities. However, they suffer from modest cycle life and relatively low power density due to slow reaction kinetics[2]. High cycle life and power density can be achieved by electric double layer capacitors (EDLCs). Based on the reversible, non-faradaic adsorption/desorption of dissolved ions on high-surface area carbonaceous electrode materials, usually activated carbon (AC) on both sides, EDLCs are capable of delivering electrical energy in a matter of seconds while granting very long cycle lives (>100 000 cycles)[3]. However, the decomposition of electrolyte moieties on the activated carbon surface restricts the cell voltage to 2.7 V at best[4,5], and the purely capacitive storage mechanism provides relatively low capacities and energy densities compared to LIBs[2].

Li-ion capacitors (LICs) can be conceived as a hybrid model bridging the gap between LIBs and EDLCs, delivering much-improved energy density compared to EDLCs while retaining similar or higher[6] power densities and having longer cycle lives compared to LIBs. LICs, in their most viable configuration, consist of a positive electrode made of AC and an intercalation-based negative electrode, typically Gr or hard carbon (HC)[7]. The negative electrode undergoes a pre-lithiation step, which provides LICs with several key advantages: Firstly, pre-lithiation eliminates the first-cycle irreversible capacity loss of the negative electrode due to SEI formation. Secondly, pre-lithiation enables the negative electrode to operate close to its lithiation potential, which not only raises the cell voltage but also minimizes its potential swing during cycling. Using negative electrodes with a low lithiation potential (~0.1 V vs. $Li/Li^+$ in the case of Gr), cell voltages of 3.8 V to 4.2 V[5,8] can be achieved. As the Gr negative electrode is typically oversized relative to the positive electrode, with common positive-to-negative (p:n) mass ratios of 1:1[9,7,6], the partial utilization of the negative electrode's capacity implies a high cycle life of LICs. Thirdly, pre-lithiation increases the LIC's capacity by enabling the AC electrode to access potentials which are negative relative to its open circuit potential (OCP), at which $Li^+$ ions are adsorbed. The extended potential range of the AC positive electrode enables the storage of about twice as much capacity compared to the positive AC electrode in EDLCs[10,11].

Despite the recent optimizations in pre-lithiation[7,12], scale-up and commercialization[8,13,14], one significant manufacturing step remains a cost-intensive parameter of cell production: The thorough drying of active and inactive materials in LIBs (and analogously, LICs) is immense and is estimated to account for 18% of manufacturing cost and 76% of energy consumption[15]. Conventionally, lithium hexafluorophosphate ($LiPF_6$) is used as the electrolyte salt in LICs. However, $LiPF_6$ readily reacts with trace water[16], producing hydrofluoric acid (HF), which not only poses safety hazards but also accelerates the degradation of numerous types of cathodes[17,18] and Gr electrodes for LIBs[19].

To address this issue and potentially reduce the drying requirements, non-hydrolyzing salts offer a promising alternative to increase water tolerance in LICs. Among these, lithium bis(fluorosulfonyl)imide (LiFSI) has emerged as one of the most viable candidates. LiFSI exhibits high chemical stability in the presence of trace water at room temperature[20,21], and its synthesis and large-scale production are becoming increasingly optimized[22]. However, the dynamics of humidity and LiFSI in LICs are mostly unknown, particularly given the fundamentally different charge storage mechanisms of AC and Gr electrodes.

A few studies exist on the impact of water-contaminated electrolytes on the performance electrochemical capacitors. Cericola *et al.* showed the aging of $ACN/TEABF_4$ in AC/AC cells with 850 ppm water, attributing degradation mostly to the negative electrode while the positive electrode was less affected[23]. However, the anion $BF_4^-$ eventually undergoes hydrolysis[24,25], and $TEABF_4$ salt cannot be considered for LICs due to the lack of $Li^+$ ions as charge carrier species. To our knowledge, only one study performed by Wang *et al.* investigated the interaction of trace water in LIC full cells based on AC and Gr[26]. However, their study uses the anion $PF_6^-$, which intrinsically limits the stability of the system due to its lability towards hydrolysis. Moreover, AC was used as the negative electrode, and Gr was the positive electrode, which does not follow the usual configuration of state-of-the-art LICs. Also, no pre-lithiation was performed on the Gr electrode despite this step being deemed a prerequisite for achieving good LIC performance[10,27].

Overall, knowledge of the impact of a water-contaminated electrolyte on the electrochemical performance of LICs is lacking. In this work, a LIC composed of an AC positive electrode and pre-lithiated Gr negative electrode is subjected to 2000 galvanostatic cycles, while the electrolyte contains defined amounts of trace water (<12 ppm, 950 ppm, 2300 ppm, and 6000 ppm $H_2O$). Also, the electrochemical stability of AC electrodes in the water-containing electrolytes is tested in AC/LFP half cells. Post-mortem analysis via SEM, EDX, and XPS yields a comprehensive overview of the relation between water contamination and the extent of chemical and morphological changes of the AC and Gr electrodes. Based on the results presented in this work, recommendations can be given on the maximum water concentration that can be tolerated in electrolytes for high-performing LICs, which is an important milestone towards minimizing the manufacturing cost allocated to maintaining dry-room conditions.

# 3 Methods

## 3.1 Materials

KS6L graphite powder (Gr, BET surface area = 20 $m^2\,g^{-1}$) was purchased from IMERYS Graphite & Carbon. YEC-8 activated carbon (AC, BET surface area = 1656 $m^2\,g^{-1}$) was purchased from Fuzhou Yihuan Carbon. Super P® conductive carbon black (CB, >99%) was purchased from Alfa Aesar. Dimethyl carbonate (DMC) and ethylene carbonate (EC) were supplied by Sigma Aldrich in battery grade levels and used without further purification. N-methylpyrrolidone (NMP) was purchased from Sigma Aldrich. Polyvinylidene difluoride (PVDF, Kynar HSV-900) was obtained from Arkema. LiFSI (99.9%, ultrapure grade) was supplied by Arkema and dried for 24 hours at 100 °C under vacuum. Double-sided carbon-primed aluminum foil (Ensafe20) was purchased from Armor Battery Group and heated at 120 °C under vacuum prior to electrode casting (see Fig. S1 for micrograph of Ensafe20 foil). Molecular sieves of 3 Å were purchased from Sigma Aldrich. Whatman GF-A separators were dried at $10^{-2}$ mbar for 48 hours at 120 °C. Lithium iron phosphate ($LiFePO_4$) electrodes (areal capacity of 2 $mAh\,cm^{-2}$) were purchased from CustomCells.

## 3.2 Electrode preparation

AC electrodes were prepared by dispersion of YEC-8 powder, CB and PVDF in NMP with a mass ratio of AC:CB:PVDF = 90:5:5 and an NMP:solid mass ratio of 3.03. The slurry was dispersed in a Retsch MM400 shaker mill for 45 min at a shaking frequency of 25 Hz and was cast onto Ensafe20 double-sided carbon-primed aluminum foil with a wet coating thickness of 150 μm. The loading of AC active material was between 3.0 to 3.3 $mg\,cm^{-2}$ and the dried electrode height was ∼80 μm. Electrodes of 18-mm and 15-mm diameters were cut from the AC cast for use in LICs and AC/LFP half cells, respectively. Gr electrodes were prepared by dispersion of Gr, CB and PVDF in NMP with a mass ratio of Gr:CB:PVDF = 90:5:5 and an NMP:solid mass ratio of 3.03 using the same shaker mill and settings. The slurry was cast onto copper foil with a wet coating thickness of 150 μm. Electrodes of 18 mm diameter were cut from the Gr cast, with a dried height of ∼80 μm and an active material loading of 3.0 and 3.3 $mg\,cm^{-2}$. $LiFePO_4$ sheets were cut into 16 mm diameter electrodes. All electrodes were dried at 120 °C and $10^{-2}$ mbar for 48 hours before transferring them into an argon-filled glove box with oxygen and water content typically <0.1 ppm.

## 3.3 Electrolyte preparation

Preparation of the dry 1 M LiFSI-based electrolyte involved the storage of a binary mixture of EC and DMC (weight ratio 1:1) over molecular sieves 3 Å (activated at 240 °C for 48 hours) to remove trace water before addition of the dried LiFSI salt. The water content of this electrolyte (hereafter referred to as LiFSI-12ppm) was <12 ppm as measured by Karl-Fischer titration. Water-containing LiFSI-based electrolytes were prepared by addition of defined amounts of MilliQ water to the LiFSI-12ppm electrolyte contained in a sealed, septum-equipped vial. Electrolytes were named according to their final water content, e.g., LiFSI-950ppm represents the EC/DMC (1:1 w/w) + 1 M LiFSI electrolyte that contains 950 ppm $H_2O$.

## 3.4 Cell assembly and electrochemical measurements

Fig. 1 shows the cell configurations, the sequence of electrochemical tests performed, and the key parameters used in the studies of LIC full cells. In step 1, the Gr electrode was electrochemically pre-lithiated in a 2-electrode Li/Gr PAT cell using 5 consecutive C/20 cycles between 0.90 V and 0.01 V with voltage hold steps of 6 h between charge/discharge steps. The sixth lithiation cycle was terminated at a capacity of 280 mAh g$^{-1}$. The electrolyte was 100 µL of LiFSI-12ppm to guarantee a high-quality solid-electrolyte interphase (SEI). The Gr electrode was retrieved from the cell, rinsed in DMC, and coupled with the AC electrode in a new 3-electrode PAT cell with Li reference ring and PEEK/gold lower plungers (positive electrode side). The injected electrolyte in the AC/Gr full cells was 100 µL of LiFSI-12ppm, LiFSI-950ppm, LiFSI-2300ppm or LiFSI-6000ppm. A mass ratio of AC:Gr = 1.00 ± 0.05 was used, based on previous studies that reported it as providing an optimal balance between power and energy density[6,9]. The cells rested at OCP for 24 hours after assembly (step 2). The LIC was initially cycled at C/10 Gr for 10 cycles (step 4, named "conditioning"), with "C" referring to the theoretical capacity of the Gr electrode (372 mAh g$_{Gr}^{-1}$). For the long-term cycling (step 5, named "cycling"), the current was increased to C/2 (charge) and 2C (discharge). Such a slow-charging regime is encouraged by a recent study[6], due to the asymmetric rate capability of Gr[28]. The LIC operates between 2.2 V and 3.8 V cell voltage, which is considered standard for Gr-based commercial LICs[6,27,29,8]. Potentiostatic EIS was conducted before and after cycling (steps 3 and 6). AC electrodes were retrieved directly from the cycled LIC, while Gr electrodes were electrochemically delithiated in a separate cell in preparation for post-mortem analysis (step 7).

Fig. 2 shows the cell configurations and electrochemical measurements performed in half cells composed of AC electrodes and Li$_{0.7}$FePO$_4$ (henceforth labeled LFP). First, LiFePO$_4$ was partially delithiated to a composition of Li$_{0.7}$FePO$_4$ in a two-electrode pouch cell against a Li counter electrode (step 1). The Li$_{0.7}$FePO$_4$ electrode was retrieved from the pouch cell, rinsed with DMC, and dried under vacuum for several minutes to remove residual solvent. It was then paired with the AC electrode in a new pouch cell, and 100 µL of the respective electrolyte was added.

After a 24-hour rest period (step 2), voltage hold tests were performed according to Weingarth *et al.*[30]. These involved charging the cell to the target voltage (e.g., 0.5 V vs. LFP, equivalent to 3.95 V vs. Li/Li$^+$) and maintaining this potential for 25 hours. Before the first hold step and following each 25-hour hold step, the cell underwent five galvanostatic charge/discharge cycles at 0.1 A g$_{AC}^{-1}$, ranging between the positive vertex potential and 0.0 V vs. LFP, to determine the capacity. This sequence of holding and cycling was repeated 16 times, resulting in a total of 400 hours at constant potential unless stated otherwise.

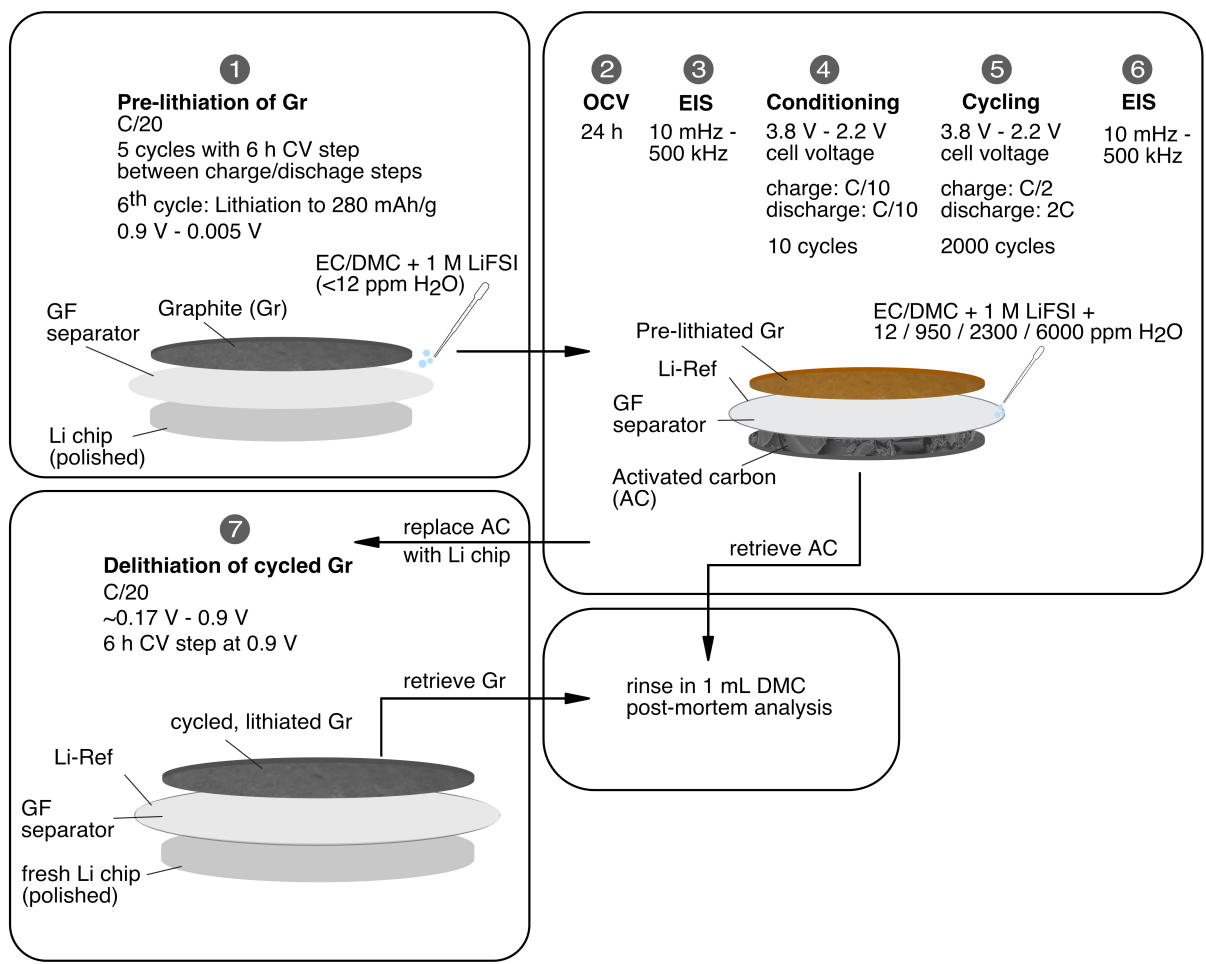

**Figure 1:** Schematic illustration of LIC cell assembly, sequence of electrochemical measurements, and preparation of samples for post-mortem analysis.

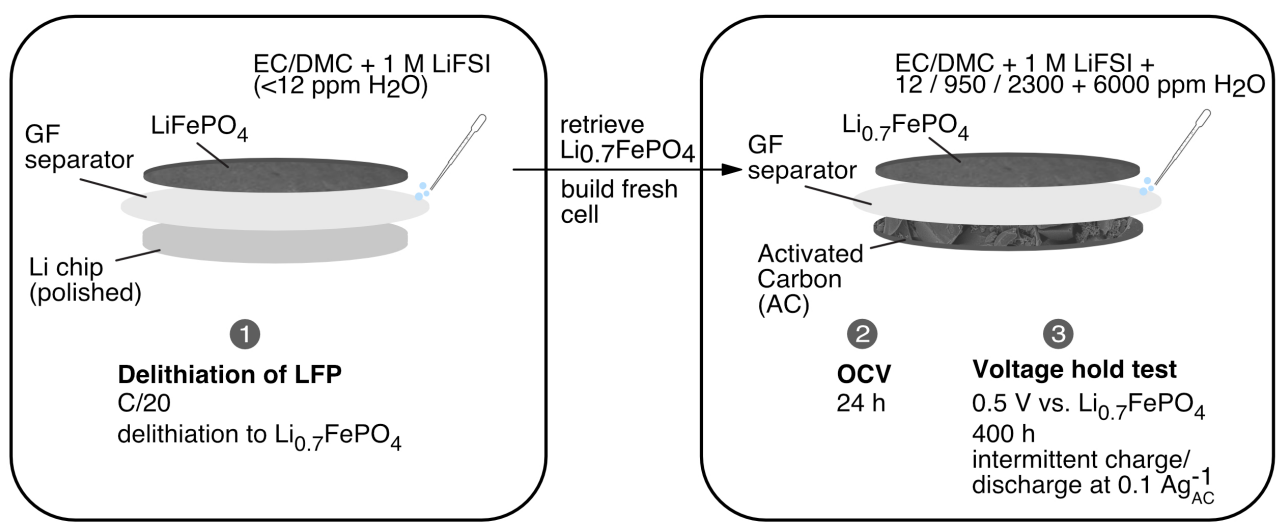

**Figure 2:** Schematic illustration of cell assembly and electrochemical measurements of AC/LFP half cells.

## 3.5 Characterization

The cycled electrodes were rinsed in 1 mL of DMC prior to post-mortem analysis. Scanning electron microscopy (SEM) was conducted using ThermoFisher Scientific Apreo SEM with T2 detector, accelerating voltages of 2-5 kV and beam current of 0.10 nA. Energy-dispersive X-ray analysis (EDX) was conducted using an Oxford Xmax 80 detector, with the accelerating voltage set to 5 kV. The samples were exposed to air for a short time during transfer from the glove box to the SEM. For X-ray photoelectron spectroscopy (XPS), samples were transferred in an argon-filled, hermetically sealed chamber from the glove box to the Kratos Analytical Axis Ultra DLD spectrometer. Monochromatic Al K$\alpha$ X-rays were used as radiation source, operating at 10 kV and 10 mA emission current. No charging compensation was used. Multiple scans were acquired for high-resolution elemental regions at a pass energy of 20 eV and averaged to enhance signal-to-noise ratio. Three spots were acquired per sample to ensure reproducibility. Since the variations among spots were low, one representative spot was selected for display. Karl-Fisher titration on electrolytes was done with Mettler Toledo C10s KF Titrator and at least 0.2 g of sample weight. $^1$H NMR spectra of the electrolytes were recorded using a Bruker BioSpin AvNeo400 spectrometer. Samples were prepared by mixing 360 µL of DMSO-d$_6$ with 40 µL of the analyte. Chemical shifts for $^1$H spectra were referenced to DMSO-d$_5$ at 2.50 ppm. Recovery of electrolytes from aged cells was carried out following the procedure shown in Fig. S2.

## 3.6 Data processing

Electrochemical data were evaluated using custom scripts in Python 3. Analysis of XPS data was carried out in CasaXPS. Spectra of cycled AC electrodes were calibrated by setting the C-C$_{sp^2}$ peak in C 1s to 284.4 eV. The spectra of cycled Gr electrodes were calibrated by setting the LiF peak in F 1s to 685.4 eV. Shirley backgrounds and symmetric Gaussian-Lorentzian peaks were used for peak deconvolution, except for graphitic carbon (C-C$_{sp^2}$), where an asymmetric peak shape was used[31]. The atomic percentage of a given component was obtained by multiplying its peak percentage with the atomic percentage of the respective element derived from the survey spectra. For the deconvolution of the S 2p regions, the 2p$_{1/2}$ and 2p$_{3/2}$ peaks of each component were constrained to match an intensity ratio of 1:2, with a peak-to-peak separation of 1.4 eV[32,33].

# 4 Results

## 4.1 Chemical compatibility of LiFSI-based electrolyte with trace water

Although it is established that LiFSI does not undergo hydrolysis at room temperature[21,20], there are limited studies investigating the chemical stability of the EC/DMC solvent when trace water is present. This investigation is crucial given that ester compounds like EC and DMC can hydrolyze under special conditions. To rule out the possibility that water reacts with the solvent at room temperature, the water content was measured using Karl-Fischer titration. Fig. 3A depicts the measured water content of LiFSI-12ppm, LiFSI-950ppm, LiFSI-2300ppm, and LiFSI-6000ppm electrolytes as a function of time after electrolyte preparation. The results show that the water amount remains constant across all samples, with minor fluctuations due to instrument error of the Karl-Fischer coulometer. This observation suggests that the water does not escape from the vial and, more importantly, is not consumed by side reactions with the solvent or the salt. To further confirm this, we present the $^1$H NMR spectrum of the LiFSI-2300ppm electrolyte after 120 days of storage at room temperature in a glove box, compared to that of the LiFSI-12ppm electrolyte (Fig. 3B). The $^1$H NMR spectra identify DMSO-$d_5$ (2.50 ppm), $H_2O$ (3.3 ppm), DMC (3.68 ppm), and EC (4.48 ppm). The $H_2O$ peak for LiFSI-12ppm likely stems from the hygroscopic DMSO-$d_6$ solvent. Minor signals indicated with an asterisk (*) correspond to methanol (3.2 ppm) and LEDC (4.10 ppm)[34]. Since these species are present in the LiFSI-12ppm electrolyte, they likely originate from impurities in EC/DMC and not from reactions of water. Notably, no new peaks appear in the LiFSI-2300ppm sample after 120 days of storage, and no peak corresponding to HF is detected (expected at ∼8.8 ppm[34]). The photographs show no color change of the water-containing electrolyte.

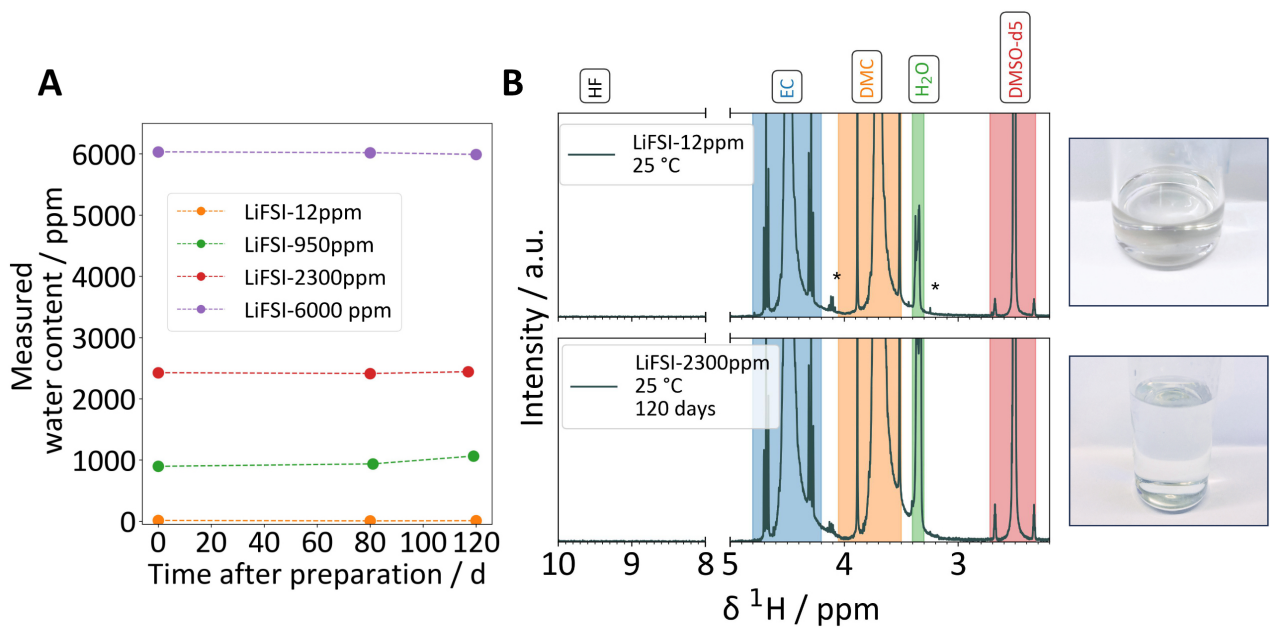

**Figure 3: A)** Water content of the prepared electrolytes as a function of storage time, measured with Karl-Fischer titration. Each electrolyte is named according to the water concentration from the first data point. **B)** $^1$H NMR (left) and photograph (right) of LiFSI-12ppm and LiFSI-2300ppm after storage at 25 °C for 120 days in a glove box.

## 4.2 Cycling of AC/Gr LICs in dry & water-containing LiFSI electrolytes

In Fig. 4, the cycling performances of the LICs cycled in LiFSI-12ppm, LiFSI-950ppm, LiFSI-2300ppm, and LiFSI-6000ppm are shown. The cycling parameters are given in Fig. 1, step 5.

The LIC cycled in LiFSI-12ppm presents a high capacity retention of 96% after 2000 cycles and high CE (Fig. 4A). The first-cycle absolute capacity of the cell is 51 $\mathrm{mAh\ g_{AC}^{-1}}$ (see Fig. S3, step 5). The OCP of the Gr electrode against $\mathrm{Li/Li^+}$ upon cell assembly (Fig. 4B) is stable at ca. 86 mV, which is characteristic for battery-type electrodes with low self-discharge[35]. During cycling, the potential swing of Gr is mostly stable, with just a mild increase in the operating potential range (against $\mathrm{Li/Li^+}$): 72 mV - 133 mV in cycles 1-3 vs. 78 mV - 160 mV in cycles 1997-2000. Such a minor increase in the potential swing means that the loss in Li inventory is minor[36]. The AC positive electrode barely changes its potential range throughout the cycling, i.e., 3877 mV - 2330 mV in cycles 1-3 vs. 3877 mV - 2358 mV in cycles 1997-2000. In total, the performance of the cell in LiFSI-12ppm is very stable and shows negligible signs of degradation.

The LIC cycled with LiFSI-950ppm exhibits a capacity retention of 96% after 2000 cycles, similar to the LIC with LiFSI-12ppm. The first-cycle capacity of 49 $\mathrm{mAh\ g_{AC}^{-1}}$ is marginally lower than that in LiFSI-12ppm. However, the occurrence of side processes related to the presence of water manifested in three aspects: Firstly, the CE is slightly lower than in LiFSI-12ppm, especially in the first cycles. This applies also to the "conditioning" cycles (see Fig. S3, step 4). Secondly, the OCP of the Gr electrode shows a slight, but steady increase to ca. 95 mV during the 24 hour resting period. This could indicate that water spontaneously reacts on the Gr electrode surface. Given the low potential of the Gr electrode, the initiation of the hydrogen evolution reaction (HER) is to be expected, likely forming LiOH on the Gr surface. Such side processes could irreversibly consume some of the lithium stored in the Gr, and subsequently increase its OCP. Thirdly, the Gr electrode operates at slightly higher potentials (155 mV - 85 mV) than in LiFSI-12ppm (133 mV - 72 mV) in cycles 1-3. This trend is continued during cycling: In cycles 1997-2000, the Gr electrode attains 108 mV (as opposed to 78 mV in the LiFSI-12ppm) at 3.8 V cell voltage and 180 mV (as opposed to 160 mV in the LiFSI-12ppm) at 2.2 V cell voltage. Thus, the higher potential of Gr for the LiFSI-950ppm electrolyte indicates some irreversible reactions that cause a loss in Li inventory and force the Gr electrode to attain a lower state of charge (SOC). Still, Gr maintains its pre-lithiated state as the inflection points in the charge-discharge curves of the Gr electrode indicate interconversion of graphite intercalation compounds. Regarding the AC positive electrode, the presence of 950 ppm $\mathrm{H_2O}$ does not appear to impact its potential swing, i.e., no significant changes compared to LiFSI-12ppm are observed.

The LIC cycled in LiFSI-2300ppm has a capacity retention just above the end-of-life (EOL) criterion of 80%, which shows that this water concentration has a clearly negative impact on the cycling stability of the LIC. Upon cell assembly, the OCP of the Gr electrode shows a distinct rise to 130 mV during the 24-hour initial resting period. This aligns with the expectation that more water causes more extensive side reactions on the lithiated Gr electrode and that more of the available Li is consumed. Due to the more pronounced loss in the Li inventory during the resting phase compared to LiFSI-950ppm, the Gr electrode's vertex potentials during cycles 1-3 are 190 mV and 102 mV, which are higher than in the case of LiFSI-950ppm. In cycles 1997-2000, the upper vertex potential of Gr has risen to 240 mV. This potential is higher than that in LiFSI-950ppm, suggesting that the loss of Li inventory during cycling is more severe than in LiFSI-950ppm. However, polarization effects, caused by the increased cell impedance (see Fig. 5), could also account for the larger potential swing of Gr in LiFSI-2300ppm. The higher cell impedance may also be the reason for the significantly lowered first-cycle absolute capacity (31 mAh $g_{AC}^{-1}$, see Fig. S3). In addition, the CE is lower than in LiFSI-12ppm and LiFSI-950ppm. Still, the degradation did not yet affect the potential swing of the AC electrode, which appeared indifferent to the cell in LiFSI-950ppm electrolyte.

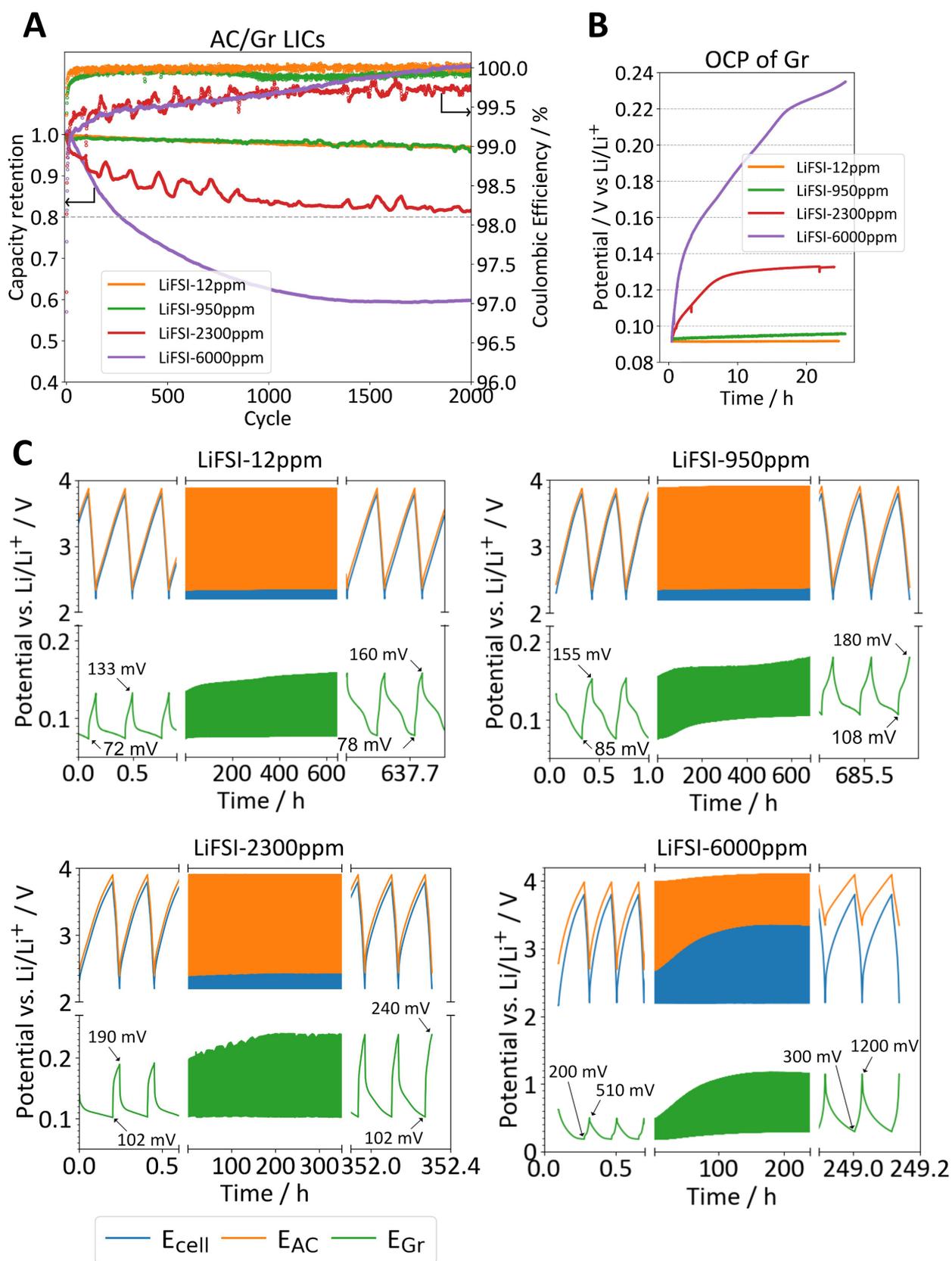

**Figure 4: A)** Capacity retention of 3-electrode LIC full cells in LiFSI-12ppm, LiFSI-950ppm, LiFSI-2300ppm and LiFSI-6000ppm electrolytes cycled between 2.2 V and 3.8 V cell voltage for 2000 cycles at a current of C/2 (charge) and 2C (discharge). **B)** OCP of Gr electrode during the 24-hour resting period after cell assembly. **C)** Voltage profiles. The first three and last three cycles are shown in a magnified view, while cycles 4-1996 are shown in a condensed view to illustrate the development of the lower and upper vertex potentials, indicated by arrows for the case of Gr.

With a capacity retention of only 62% after 2000 cycles and the lowest first cycle absolute capacity (28 mAh $g_{AC}^{-1}$, see Fig. S3), the cell cycled in LiFSI-6000ppm shows the poorest performance. The unstable cell performance is already indicated by the development of Gr's OCP upon assembly. The OCP presents a large drift, from 90 mV to 230 mV vs. $Li/Li^+$ during 24 hours. This indicates that water initiates severe side reactions. The potential swing of Gr during cycling demonstrates that the water contamination leads to a drastic loss of Li inventory: In cycles 1-3, Gr reaches a potential of only 200 mV when the LIC is fully charged. This implies that Gr attains a low state of charge, considering that significant Gr lithiation typically occurs below 200 mV[37]. However, it should be noted that for LiFSI-6000ppm, a significant increase in the impedance is observed for the Gr electrode (see Fig. 5), which complicates the determination of the exact lithium content of Gr. Nevertheless, it is safe to say that the already limited Li inventory in the cell is further decimated during cycling, as shown by the steady drift in the Gr electrode's operating potential, which alternates between 300 mV and 1200 mV in cycles 1997-2000. Notably, and contrary to the other systems with lower amounts of water contamination, the AC electrode's potential swing is significantly affected: The potential swing of the AC positive electrode changes from 3982 mV - 2860 mV in cycles 1-3 to 4093 mV - 3445 mV in cycles 1997-2000. The increase of the AC upper cutoff potential by ca. 220 mV compared to the AC in LiFSI-12ppm (3877 mV) increases the propensity of electrolyte decomposition, provoking capacity fade on the AC side[4,5]. On the other hand, the development of the AC's lower vertex potential is in agreement with the pronounced Li loss over the course of cycling: In the first cycles, AC's lower vertex potential extends below its potential of zero charge (pzc, ca. 3150 mV vs. $Li/Li^+$ [4,5]), indicating that, initially, $Li^+$ ions can adsorb on its surface upon full discharge of the LIC. These $Li^+$ ions are supplied by the pre-lithiated negative electrode[38,28]. However, after ca. 1000 cycles, AC is cycling strictly above its pzc, manifesting the inability of the Gr electrode to provide enough $Li^+$ ions for adsorption during LIC discharge. The voltage profile of the LIC after ca. 1000 cycles, (i.e., AC not obtaining potential values below its pzc) is similar to that of a non-pre-lithiated LIC[36,10]. Therefore, the water contamination removes the benefits of the Gr pre-lithiation step. Aside from this, the prevalence of parasitic processes in LiFSI-6000ppm is also illustrated by the coulombic efficiency, which is significantly lower than in the cells with less water contamination, especially in the first 1000 cycles where most of the capacity fade occurs.

In total, the electrochemical cycling data showed that increasing the electrolyte's water content accentuates degradation effects. These effects are evident through lower coulombic efficiency, reduced absolute capacity, and capacity retention, and a positive shift in the operating potential of the Gr electrode. However, at 950 ppm water content, the degradation effects are rather subtle, allowing for a remarkably good performance similar to that of the dry electrolyte. In contrast, higher water contents of 2300 ppm and 6000 ppm display more pronounced signs of degradation.

## 4.3 Electrochemical Impedance Spectroscopy

To get more complementary information on cell degradation, EIS was performed. In Fig. 5, the Nyquist plots acquired at step 3 (before cycling) and step 6 (after 2000 cycles) are presented for the LICs in LiFSI-12ppm, LiFSI-950ppm, LiFSI-2300ppm, and LiFSI-6000ppm. The left and middle column represent the Nyquist plot on the AC side and Gr side, respectively. The right column represents the impedance of the LIC full cell, i.e., measured between AC and Gr electrode.

In LiFSI-12ppm, the Nyquist plots on the AC side are identical with only a marginal increase observed in the electrolyte resistance, identified by the high-frequency intercept of the semi-circle on the x-axis. Notably, the Nyquist plots have a quasi-vertical branch at low frequencies, indicative of ideal capacitive behavior[39]. The Gr impedance spectra show negligible changes as well, resulting in similar LIC full-cell spectra that are only slightly shifted due to the marginal increase in electrolyte resistance. The marginal increase in resistance is consistent with the cycling results where a negligible capacity fade is observed, and invariably, minimal electrolyte decomposition.

In LiFSI-950ppm the Nyquist plot on the Gr side at step 3 shows an increase in the diameter of the two semicircles and a higher total resistance compared to LiFSI-12ppm. This is in line with the OCP increase during resting, and indicates that the SEI of Gr is transformed upon exposure to the water-contaminated electrolyte. Interestingly, the Nyquist plot on the Gr side shows a slight decrease in the diameters of the two semi-circles and a slightly lower total resistance after 2000 cycles. This may indicate that Gr sustains little damage during cycling, and is in line with the overall good capacity retention. Similar observations were reported by Burns *et al.*[40], where the beneficial impact of moisture is recorded in LIB using electrolytes containing 1000 ppm of $H_2O$. However, the Nyquist plots on the AC side are more inclined in the low-frequency domain compared to LiFSI-12ppm, and cycling appears to further decrease the slope of the low-frequency branch. This signifies that the AC starts losing its ideal capacitive properties when the LIC is exposed to LiFSI-950ppm. Thus, the EIS data suggest that aging is more prominent on the AC side than on the Gr side if the water level is at 950 ppm.

When the water content in the electrolyte is increased to 2300 ppm, the total resistance on the AC side is increased and the low-frequency capacitive branch is further inclined, representing a further departure from ideal capacitive behavior. The Gr electrodes' total resistance before cycling is higher than in LiFSI-12ppm and LiFSI-950ppm, suggesting that exposure to 2300 ppm water causes extensive irreversible interfacial changes, which is in line with the rise in OCP during the 24-hour resting period. Cycling further increases the total resistance of the Gr electrode. Therefore, it is probable that a transition point could exist between the minimal effect at 950 ppm and detrimental effect beyond 2300 ppm as mentioned earlier. This is further supported by the impedance in LiFSI-6000ppm, where the AC electrode displays a significantly increased total resistance after 2000 cycles and a pronounced 45° inclined line at medium-high frequencies, which is indicative of increased pore diffusion resistance[41,39]. The increased pore diffusion resistance likely stems from extensive electrolyte decomposition and pore-blocking. The unstable cycling behavior of Gr is reflected by a very high total resistance after 2000 cycles.

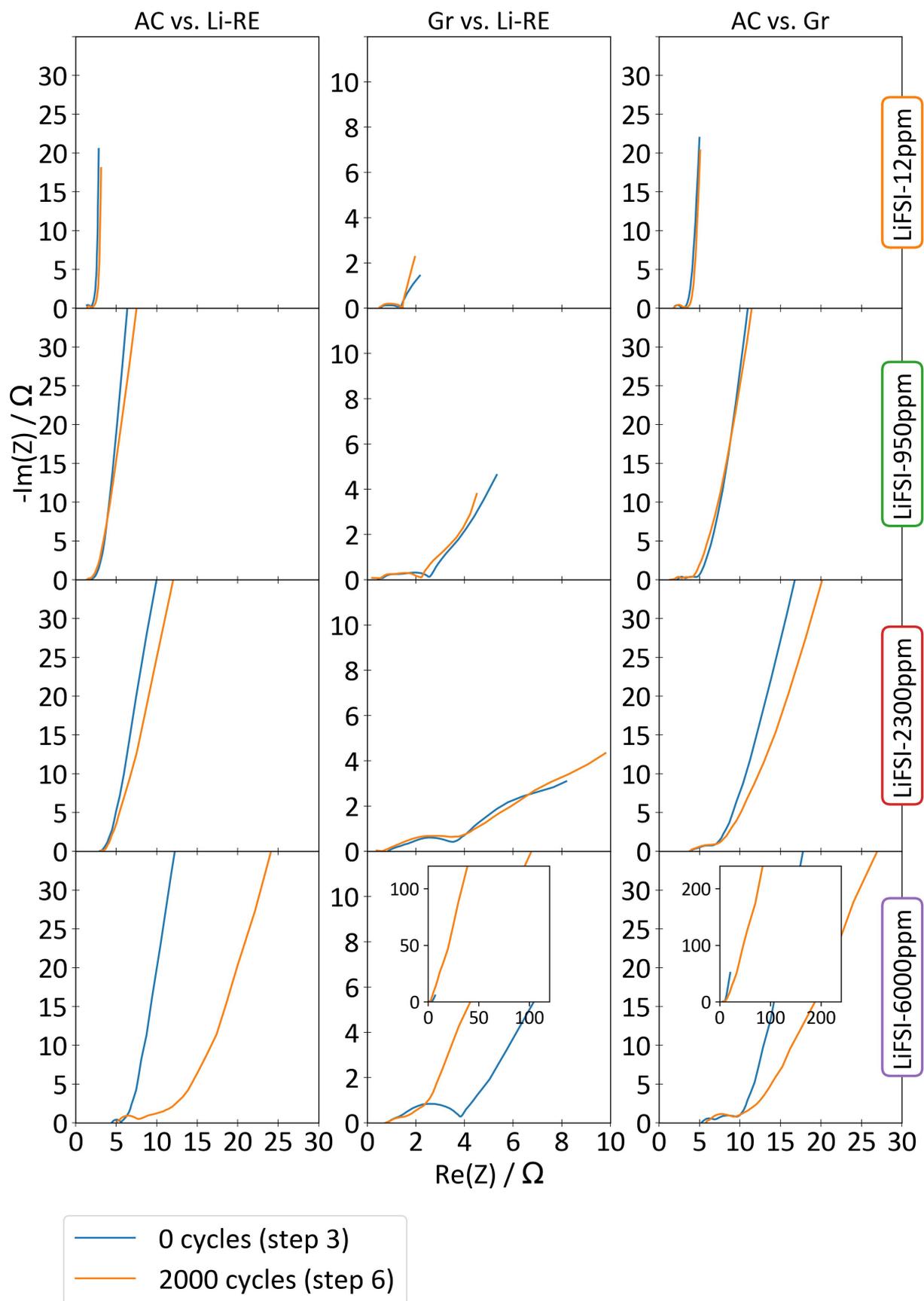

**Figure 5:** EIS for LICs after 24-hour OCP (blue, step 3) and after 2000 cycles (orange, step 6) in LiFSI-12ppm, LiFSI-950ppm, LiFSI-2300ppm, and LiFSI-6000ppm measured in three-electrode setup: AC side (left), Gr side (middle), AC vs. Gr (right). Selected insets show the complete Nyquist plots where the low-frequency section extends beyond the borders of the main plots.

Overall, the EIS spectra align with cycling results, indicating minimal aging and potentially even a minor positive effect on the Gr electrode at water concentrations up to 950 ppm. Water concentrations of 2300 ppm and 6000 ppm have a clear detrimental effect, and lead to severe transformation of the electrode/electrolyte interface due to the very high moisture content.

### 4.4 Post-mortem analysis of AC positive electrodes

The AC positive electrodes were retrieved from the cycled LICs for post-mortem analysis. The nomenclature of the AC electrodes contains the amount of water contamination in the electrolyte, e.g. AC-950ppm denotes the AC electrode retrieved from the LIC cycled in the electrolyte LiFSI-950ppm. AC-pristine denotes the uncycled AC electrode.

#### 4.4.1 Post-mortem SEM of AC

Fig. 6 shows electron micrographs of AC-pristine, AC-12ppm, AC-950ppm, AC-2300ppm, and AC-6000ppm.

AC-pristine consists of irregularly shaped, micron-sized AC particles interspersed with finer, granular conductive carbon particles sized ∼50 nm. The surface of AC particles looks mostly smooth, while some roughness indicates the micro- and mesopores that cannot individually be resolved by SEM (Fig. 6B).

AC-12ppm shows no striking morphological differences compared to the AC-pristine when observed at low magnifications. However, at higher magnifications (Fig. 6D), certain areas of the electrode surface exhibit smooth, film-like, and partly spherical deposition structures. These could potentially originate from the decomposition of the carbonate solvent and/or LiFSI salt.

AC-950ppm exhibits spherical features scattered throughout the electrode surface. Fig. 6G shows that the deposits are partly interconnected and agglomerated. EDX mapping shows that the deposits mainly consist of F and O (see Fig. S4A). This might indicate that water favors the oxidative decomposition of LiFSI salt. The DFT study of Di Muzio *et al.* identifies LiF as a hydrolysis product of LiFSI, but with a relatively high activation barrier of 63 kJ mol$^{-1}$, preventing the reaction at room temperature[21]. However, this energy could potentially be supplied electrochemically within a cell, enabling hydrolysis — and thus LiF generation. In addition to the spherical F-rich deposits, a few selected spots of the electrode exhibit fine and grainy deposits of a few nm in size (Fig. 6F). Such features are not observed in the electrode cycled in LiFSI-12ppm, and could indicate partial clogging of the pores of AC. However, the high capacity retention of 96% after 2000 cycles suggests that neither the spherical deposits nor the grainy, nm-sized deposits infiltrate into AC's micro- and mesoporous network and that they instead cover the external electrode surface only. Considering the AC's surface area of ∼1500 m$^2$ g$^{-1}$, the deposits on the external surface only affect a rather small fraction of the total electrode surface available for ion adsorption. The low effect of the deposits on the electrochemical integrity of AC is also supported by EIS, which only shows a slight inclination of the low-frequency, capacitive branch, but no other changes (see Fig. 5).

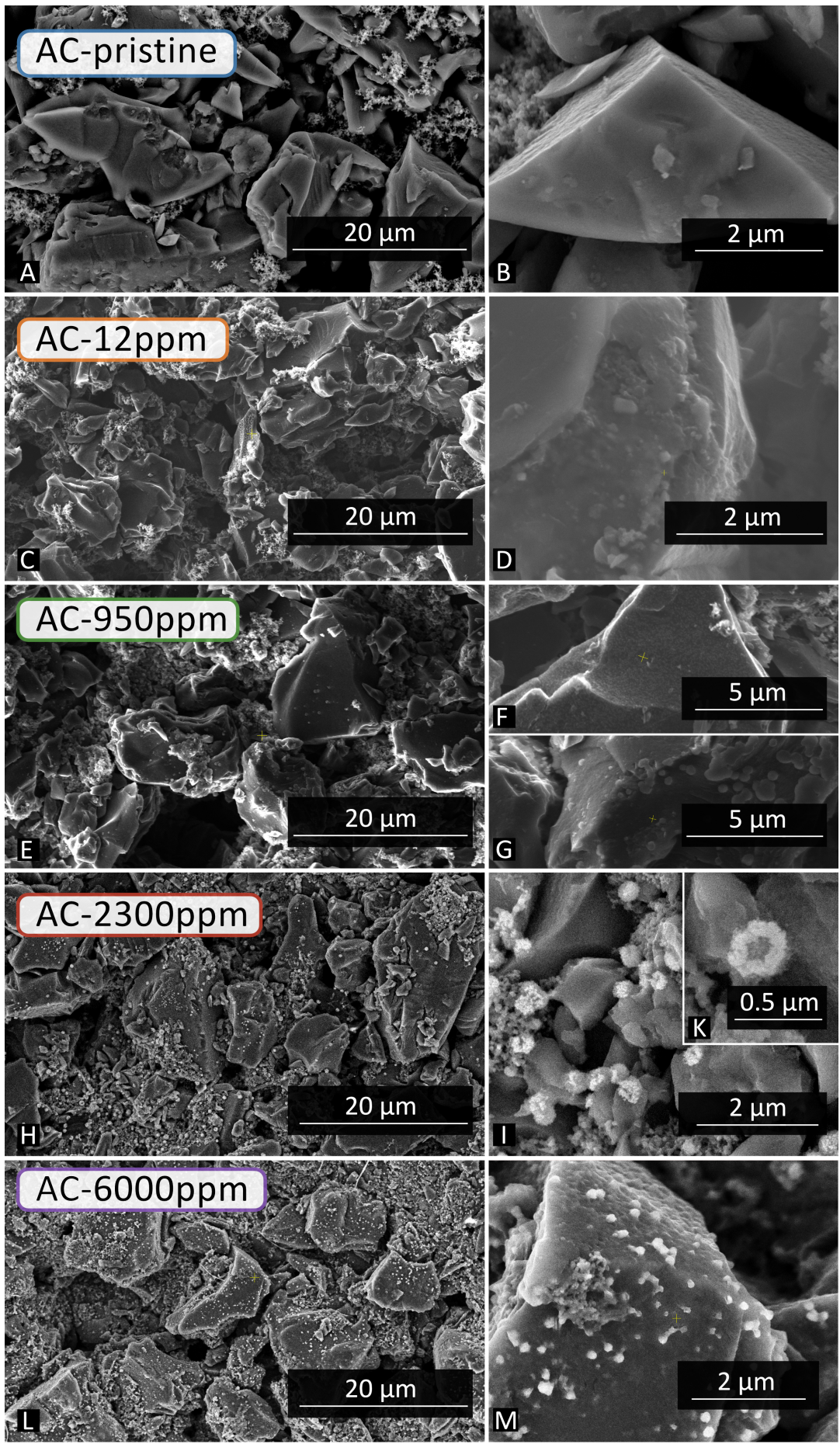

**Figure 6:** SEM micrographs of **A-B)** AC-pristine, **C-D)** AC-12ppm, **E-G)** AC-950ppm, **H-K)** AC-2300ppm, **L-M)** AC-6000ppm.

The low-magnification view of AC-2300ppm (Fig. 6H) reveals more abundant surface deposits compared to AC-950 ppm. The high-magnification view (Fig. 6I and K) shows that the spherical deposits have a coarse, frayed texture, and EDX mapping (Fig. S4B) identifies F, S, and O as their main constituents. The similar composition of these deposits to those in AC-950ppm, but their greater abundance, suggests that increased water content exacerbates LiFSI decomposition. However, the degradation, migration or agglomeration of PVDF binder could also be a possible source for F. It is known that hydroxide, plausibly originating from the contact of water with the lithiated Gr surface, can decompose PVDF[42]. However, the concentration of $OH^-$ is presumably not high enough in the present system, since PVDF possesses excellent stability in aqueous hydroxide solutions up to 20%[43]. Additionally, the C 1s and F 1s spectra (see section 4.4.2) show no evident signs of PVDF degradation, therefore the deposits are more likely to stem from salt decomposition rather than binder degradation. Nevertheless, the densely distributed deposits on AC-2300ppm illustrate that water accelerates aging processes of AC, which is well in line with the increased total resistance and lower slope of the capacitive branch of the AC-2300ppm electrode compared to AC-950ppm.

The surface of AC-6000ppm (Fig. 6L and M) is widely covered by spherical deposits, and their prevalence seems higher than in AC-2300ppm. In addition, a surface film can be clearly recognized, which presumably stems from carbonate decomposition and polymerization. This suggests significant pore blockage, causing restricted ion diffusion, which agrees with the pronounced 45° segment and the high total resistance in the Nyquist plot of the AC-6000ppm electrode (see Fig. 5).

### 4.4.2 Post-mortem XPS of AC

Fig. 7 displays the atomic percentages obtained from XPS survey spectra of AC-pristine, AC-12ppm, AC-950ppm, and AC-2300ppm.

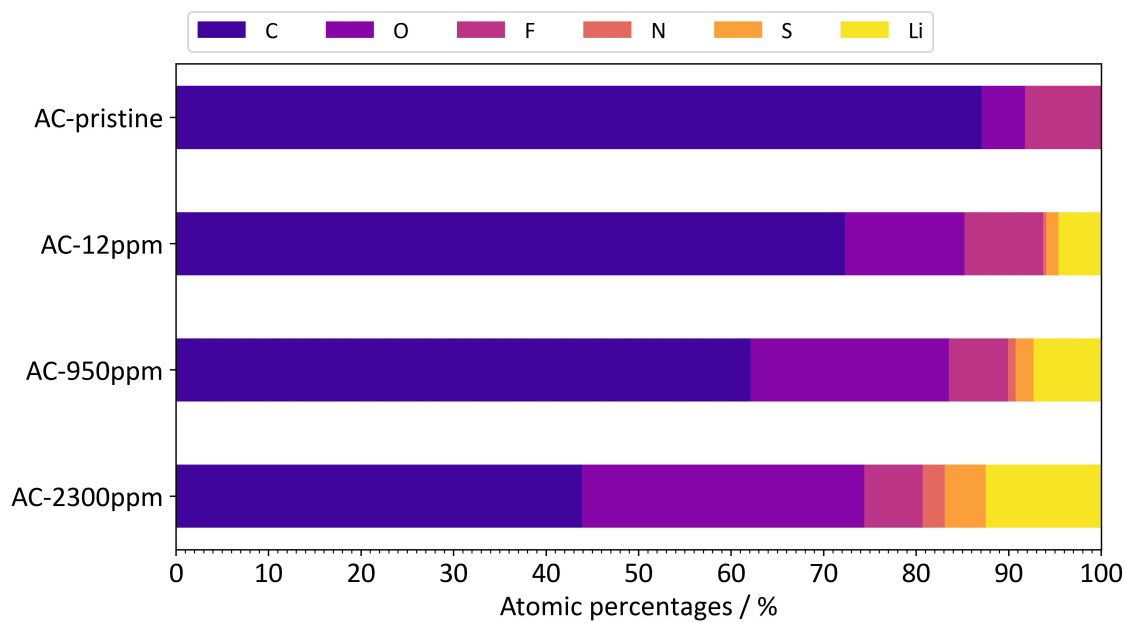

**Figure 7:** Atomic percentages of C, O, F, N, S and Li extracted from survey spectra of AC-pristine, AC-12ppm, AC-950ppm and AC-2300ppm. Raw survey spectra can be found in Fig. S6.

In AC-pristine, carbon (C) constitutes 80 at% from the AC and conductive additive, with oxygen (O) originating from their surface groups and fluorine (F) from the PVDF binder. A comparison between AC-pristine and AC-12ppm shows an increase in oxygen content, suggesting that carbonate solvent decomposition and deposition occurs in LiFSI-12ppm, i.e., in nearly dry conditions. This is in line with a study of Nyholm *et al.* which proposes that the carbonate solvent decomposition initiates near 3.3 V vs. $Li/Li^+$ [44]. Alternatively, the increased O content might stem from oxidation of the carbon surface, as proposed in our previous study[4] and by a study of Song *et al.* [45]. The at% of O continues to rise in AC-950ppm and AC-2300ppm. This suggests that the hydrolysis of the carbonate solvent, which was found not to occur during storage of the electrolyte at room temperature (see Fig. 3), is facilitated on the positive AC electrode. Simultaneously, the amounts of lithium (Li) increase, which may be attributed to the formation of lithium-containing organic deposits or the decomposition of the LiFSI salt. The rising amounts of S and N from AC-12ppm to AC-2300ppm are plausible indicators that the LiFSI salt is undergoing decomposition as the water amount is increased. Notably, no aluminum was detected in any survey spectrum, which would otherwise indicate dissolution and redeposition due to current collector corrosion. Severe aluminum corrosion has been observed in AC/AC EDLCs operating in water-contaminated TFSI-based electrolytes[46], and LiFSI in EC/DMC corrodes aluminum at potentials above 4.15 V vs. $Li/Li^+$ [20,4]. Trace water could potentially lower this threshold and potentially cause corrosion near the AC's positive vertex potential in the LICs of this study (∼3.95 V vs. $Li/Li^+$). However, visual inspection of the cycled AC electrodes revealed no degradation, suggesting that the Ensafe20 current collector's carbon priming layer effectively protects the aluminum foil (see Fig. S1).

Fig. 8 shows the high-resolution C 1s, F 1s, S 2p, and N 1s spectra of AC-pristine, AC-12ppm, AC-950ppm, and AC-2300ppm.

In the C 1s spectrum of AC-pristine, graphitic carbon (C-C$_{sp^2}$) is the most dominant species. The corresponding peak at 284.4 eV is modeled with an asymmetric peak shape[31] and the atomic percentage of this component is 54 at%. The peak at 285.5 eV corresponds to sp$^3$-type carbon[47] or adventitious carbon (C-H). Peaks at binding energies are attributed to oxygen-terminated surface groups that typically result from the activation process of AC[47]. Oxygen-bound carbon, summarized as C-O/C=O/O-C=O in the atomic percentage diagram, has an abundance of ca. 15 at%. The peak at 290.5 eV and the peak at 688 eV in the F 1s spectrum originate from $CF_2$ of the PVDF binder[48]. The S 2p and N 1s spectra of AC-pristine are featureless and therefore not shown. Instead, the S 2p and N 1s spectra of AC-pristine dipped in LiFSI-12ppm without subsequent rinsing is shown (marked with * in Fig. 8) to represent the chemical signature of the $FSI^-$ molecule adsorbed on the AC surface.

The C 1s spectra reveal that the at% of C-C$_{sp^2}$ progressively decreases from AC-pristine to AC-12ppm, and further to AC-950ppm and AC-2300ppm. This trend suggests the formation of surface deposits and that this surface film grows in thickness with increasing water content, in agreement with the observation in SEM. While the SEM images did not allow an unambiguous statement regarding the formation of surface film on AC-12ppm, the decline of C-C$_{sp^2}$/C-H species indicates that slight solvent decomposition and deposition already occurs in AC-12ppm. Concomitantly, the atomic fractions of C-C$_{sp^3}$ and oxygen-bound carbon (summed up as C=O/O-C=O/C-O) increase for increasing water concentration. This illustrates that the surface layer consists of aliphatic hydrocarbons with oxygen-containing organic moieties, and indicates that the carbonate solvent decomposes more extensively with increasing water content.

In the F 1s spectra, the atomic percentage of assigned to $CF_2$ groups of PVDF decreases progressively. This trend is consistent with the increasing coverage of the electrode by surface products, forming an interlayer that attenuates the binder signal. However, the decrease in the $CF_2$ signal is negligible between AC-pristine and AC-12ppm, with the atomic percentages remaining nearly equal. This observation aligns well with SEM results, which did not reveal a clearly visible surface layer for the AC-12ppm sample. A small amount of LiF is observed in AC-12ppm, indicating that anion decomposition is minimal in the LiFSI-12ppm electrolyte. However, in the AC-950ppm sample, there is a significant increase in the amount of LiF. Since LiF is a byproduct of LiFSI hydrolysis[21], this suggests that higher water amounts promote LiFSI hydrolysis under electrochemical polarization. The F-rich spherical deposits observed in SEM-EDX for the AC-950ppm sample (Fig. 6) are therefore primarily composed of LiF. Interestingly, in the AC-2300ppm sample, a new component emerges at 691 eV. This peak could suggest the formation of multi-fluorinated carbon[49]. However, no corresponding peaks for $CF_3$ or $CF_x$ (expected at ∼293 eV) are present in the C 1s spectra, which challenges this interpretation. Yazami et al.[50] attributed the peak at 691 eV to O-F bonds; however, the absence of a corresponding feature in the O 1s spectra in our work (see Fig. S7) does not confirm the presence of O-F bonds. Therefore, the origin of this signal remains unclear and is designated as unknown component.

In the S 2p spectrum of AC-12ppm, a new component (blue doublet) appears alongside the pristine LiFSI salt (green doublet). A similar component was observed on AC electrodes subjected to voltage hold tests at 3.95 V vs. Li/Li$^+$ in the same electrolyte[4], indicating that this component also forms under galvanostatic cycling conditions. For AC-950ppm and AC-2300ppm, satisfactory peak fitting was only achieved by introducing an additional oxidation product (yellow doublet). This suggests that water triggers the hydrolysis of LiFSI towards the formation of various oxidation products. Notably, the atomic concentrations of LiFSI-ox-1 and LiFSI-ox-2 are the highest for AC-2300ppm.

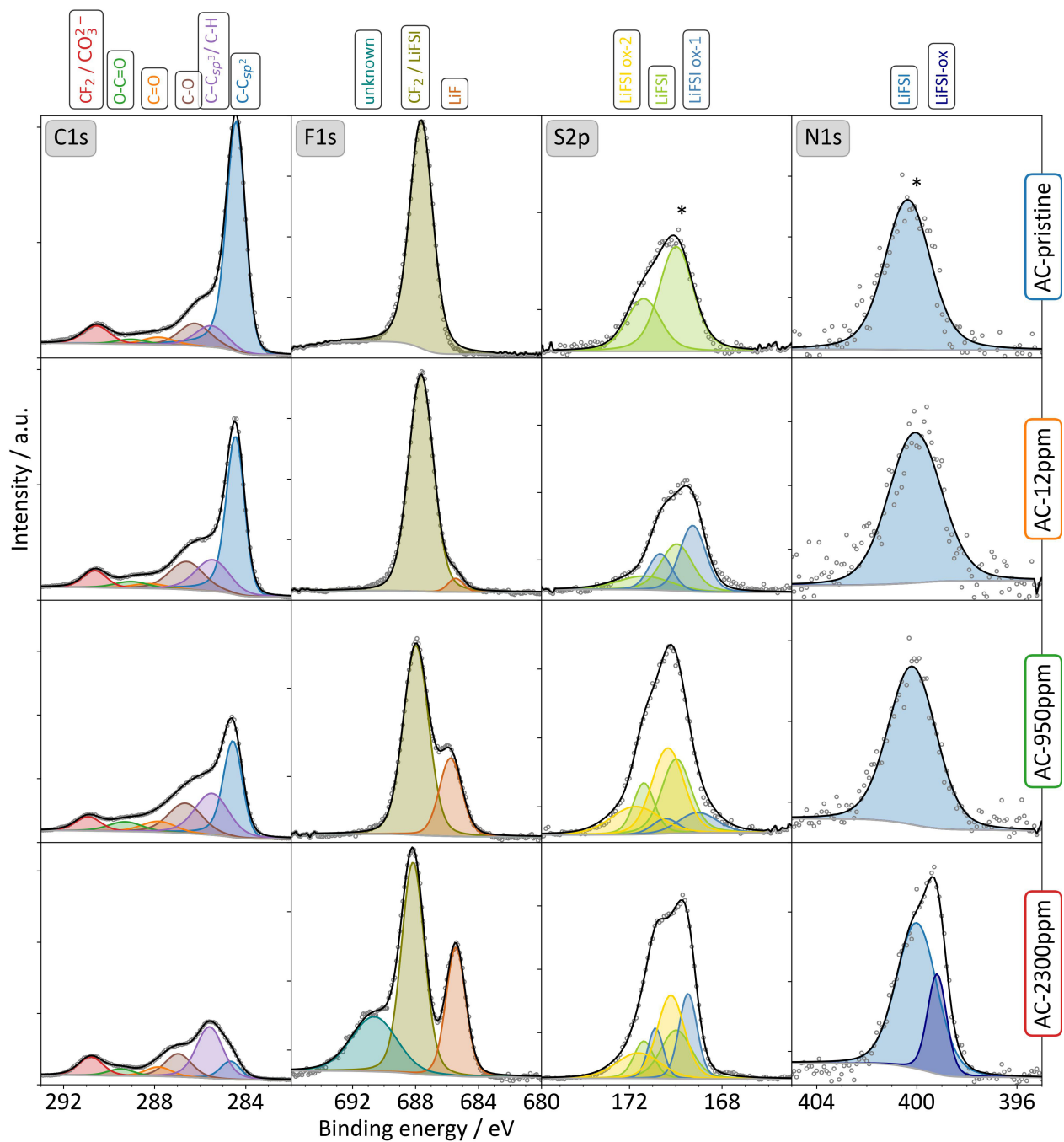

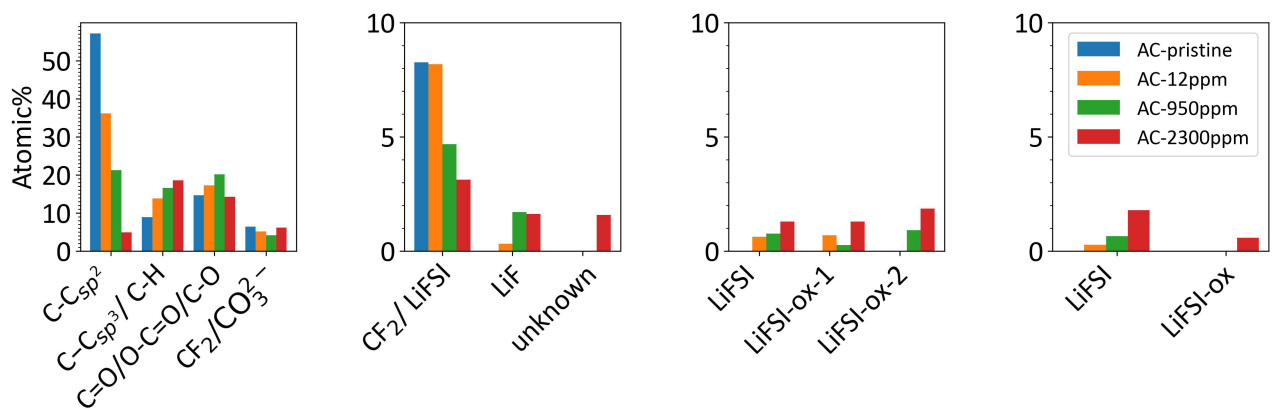

**Figure 8: A)** Post-mortem XPS of AC electrodes: C 1s, F 1s, S 2p, and N 1s regions (O 1s and Li 1s regions are contained in Fig. S7). Spectra are scaled arbitrarily and calibrated by setting C-C$_{sp^2}$ to 284.40 eV. **B)** Atomic percentages of components.

The N 1s spectra primarily show the signal of the pristine LiFSI salt adsorbed on AC, with the exception of AC-2300ppm, which exhibits an additional component at 399 eV. This new peak is likely due to the oxidation of LiFSI, which is in good agreement with the decomposition products observed in the S 2p spectrum and the high amount of LiF in AC-2300ppm. In contrast, no distinct peak for LiFSI oxidation products is observed in the N 1s spectra of AC-950ppm and AC-12ppm. This suggests that, in these cases, anion oxidation may not proceed through S-N bond cleavage, or that the oxidation products have a chemical shift indistinguishable from that of the pristine salt.

In summary, the combined XPS and SEM analysis of the AC positive electrodes unveiled that increasing amounts of water contamination increase the extent of carbonate decomposition and formation of a CEI-like surface film, especially in AC-2300ppm. A possible cause might be the electrolysis of water, resulting in hydroxide ions on the Gr side and protons on the AC side, both of which promote the hydrolysis of the ester functionalities of EC and DMC. The S 2p, N 1s and F 1s spectra illustrated that the LiFSI anion appears to undergo hydrolysis under electrochemical polarization. In particular, the anion decomposition at a water concentration of 2300 ppm leads to a widespread formation of $Li_xO_yF_z$ clusters.

## 4.5 Post-mortem analysis of Gr negative electrodes

The Gr electrodes were electrochemically delithiated prior to post-mortem analysis by replacement of the AC positive electrode with a Li chip and discharge to 0.9 V vs. Li/Li$^+$ (see Fig. 1, step 7). Gr-12ppm denotes the Gr electrode retrieved from the LIC cycled in LiFSI-12ppm. The same nomenclature applies to Gr-950ppm, Gr-2300ppm, and Gr-6000ppm. Gr-pristine denotes the uncycled Gr electrode.

### 4.5.1 Post-mortem SEM of Gr

Fig. 9 shows electron micrographs of Gr-pristine, Gr-12ppm, Gr-950ppm, Gr-2300ppm, and Gr-6000ppm. Gr-pristine (Fig. 9A and B) consists of micron-sized, flake/platelet-shaped Gr particles with clearly visible edges and well-defined particle boundaries. The conductive carbon additive is uniformly distributed between the Gr flakes.

Gr-12ppm (Fig. 9C and D) shows a smoother, less textured surface than the pristine Gr electrode. The previously well-defined, relatively sharp boundaries of the Gr flakes are now obscured by a thin film. This film likely formed due to the reductive decomposition of the carbonate solvent and LiFSI (reduced below 1.9 V vs. Li/Li$^+$ [49]) during the formation cycles of the electrochemical pre-lithiation step (step 1 in Fig. 1), creating an SEI layer. Additionally, Gr-12ppm exhibits spherical deposits with diameters ≤0.3 μm. Similar deposits were reported for Gr subjected to three C/10 formation cycles in 1 M LiFSI in DMC/adiponitrile[51]. Andersson *et al.* identify such deposits as crystalline LiF[52], albeit using a $LiPF_6$-based EC/DMC electrolyte. Zhang *et al.* suggest the creation of such spherical deposits is promoted at high currents[53], suggesting that they are preferentially formed during the cycling of the LIC rather than during the Gr formation/pre-lithiation step. The EDX-mapping in Fig. S5A confirms that the spherical deposits are rich in F, thus suggesting that their main component is LiF. We rule out the possibility of Li plating as a cause for the spherical deposits, as the Gr potential remained well over 0 V vs. $Li/Li^+$ during cycling of the LIC (see Fig. 4C), and a slow-charging regime was adopted (see step 5 in Fig. 1) as it is reported to suppress Li plating[6].

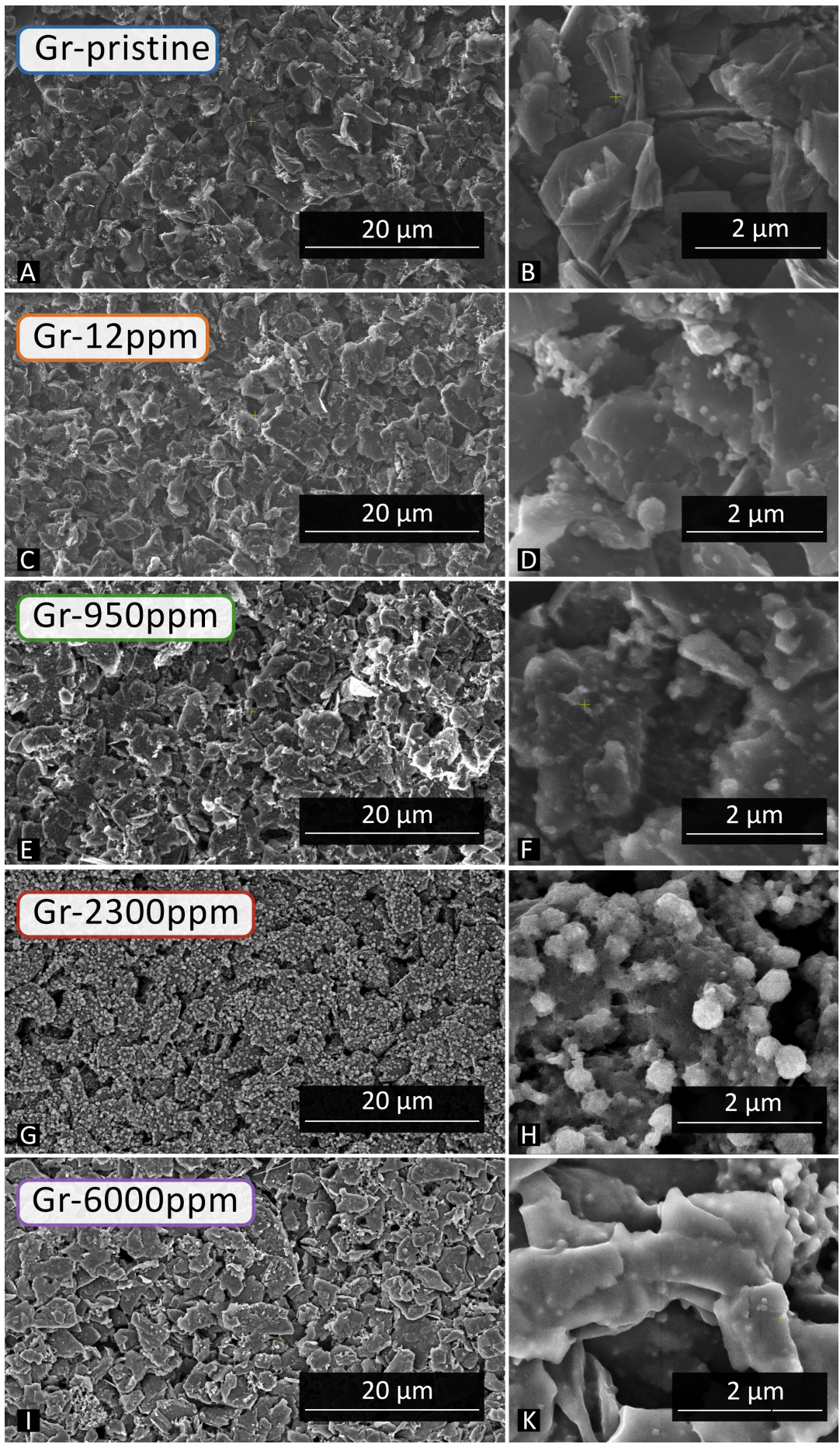

**Figure 9:** SEM micrographs of **A-B)** Gr-pristine, **C-D)** Gr-12ppm, **E-F)** Gr-950ppm, **G-H)** Gr-2300ppm, **I-K)** Gr-6000ppm.

Spherical deposits of similar size, morphology and quantity are observed on Gr-950ppm (Fig. 9E and F). EDX mapping (see Fig. S5B) confirms that the spherical deposits have a similar composition as in Gr-12ppm, i.e., an enrichment in F and O. In addition to the spherical deposits, the surface of the Gr particles in Gr-950ppm appears smoother and less defined compared to Gr-12ppm. This suggests that an increase of water contamination from <12 ppm to 950 ppm causes a thickening of the SEI, likely due to promoted solvent decomposition, which results in polymeric/oligomeric SEI products. Plausibly, water facilitates the ring-opening hydrolysis of ethylene carbonate and its subsequent polymerization reactions[54].

While the morphological differences between Gr-12ppm and Gr-950ppm are rather subtle, significant changes occur as the water concentration is increased to 2300 ppm (Fig. 9G and H). The Gr-2300ppm electrode becomes densely covered with spherical deposits, which are significantly more abundant than those on Gr-12ppm and Gr-950ppm. These deposits grow up to 500 nm in size, and EDX mapping (Fig. S5C) reveals that the main constituents of these deposits are F and O, with a notable presence of S, suggesting the inclusion of LiFSI fragments. Additionally, the Gr particles are coated with a much thicker, interconnected SEI film compared Gr-950ppm. This suggests that the increase of water concentration to 2300 ppm further facilitates the decomposition and polymerization reactions of EC and DMC, presumably catalyzed by the generation of $OH^-$ on Gr side or $H^+$ on AC side. The extensive surface coverage on Gr-2300ppm corresponds with the observed increase in the Gr's total resistance after 2000 cycles (see Fig. 5). The SEI has grown even thicker on Gr-6000ppm (Fig. 9I and K). The resistive nature of the SEI is indicated by the observed high total resistance on the Gr side (see Fig. 5).

### 4.5.2 Post-mortem XPS of Gr

Fig. 10 displays the atomic percentages obtained from XPS survey spectra of Gr-pristine, Gr-12ppm, Gr-950ppm, and Gr-2300ppm.

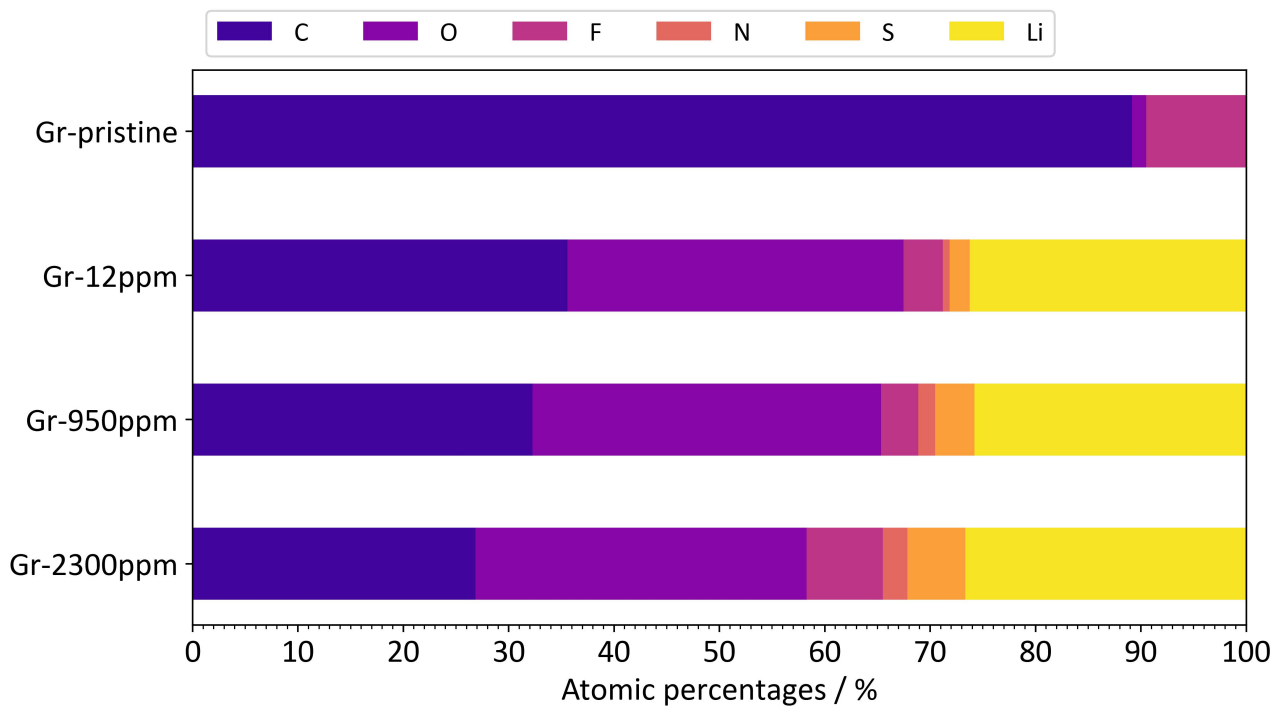

**Figure 10:** Atomic percentages of C, O, F, N, S, and Li extracted from survey spectra of Gr-pristine, Gr-12ppm, Gr-950ppm, and Gr-2300ppm. Raw survey spectra can be found in Fig. S6.

The XPS survey reveals distinct trends in the atomic percentages of elements for Gr negative electrodes as the water amount in the electrolyte is increased. In Gr-pristine, the SEI is composed primarily of C, with minor contributions from O due to surface groups of Gr and conductive additive, and F originating from the PVDF binder. In Gr-12ppm, the significant increase in O indicates the reductive decomposition of the carbonate solvents EC and DMC associated with SEI formation, along with a notable presence of Li in the SEI components. As the water content increases from <12 ppm to 950 ppm and then to 2300 ppm, the atomic fraction of C declines, suggesting the formation of more SEI products and a thicker SEI consistent with the SEM observations of a very thick morphology. Concurrently, the fractions of S, N, and F increase, suggesting that higher water content promotes the reductive decomposition of LiFSI and the integration of its fragments into the SEI.

Fig. 11 shows the high-resolution C 1s, F 1s and S 2p spectra of Gr-pristine, Gr-12ppm, Gr-950ppm, and Gr-2300ppm. The XPS spectra of Gr-pristine are discussed before the trends within the respective high-resolution spectra are addressed.

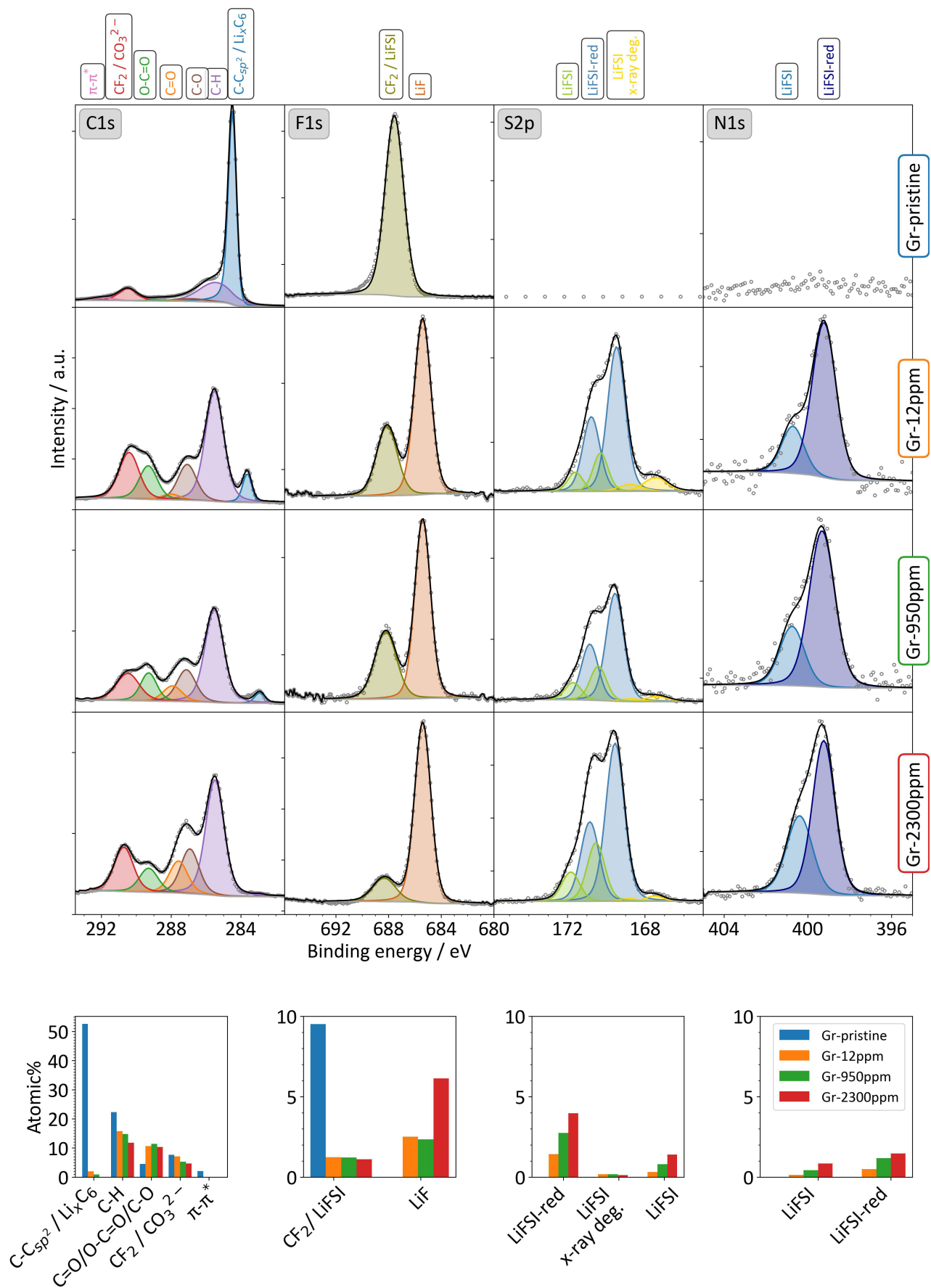

**Figure 11: A)** Post-mortem XPS of Gr electrodes: C 1s, F 1s, S 2p and N 1s regions (O 1s and Li 1s regions are contained in SI). Spectra are scaled arbitrarily and spectra of cycled Gr electrodes are calibrated by setting LiF to 685.40 eV. **B)** Atomic percentages of components.

The C 1s spectrum of Gr-pristine is similar to that of AC-pristine, with the exception that the at% of oxygenated carbon surface groups (C-O/C=O/O-C=O) only constitutes 5 at% as opposed to 15 at% in AC-pristine. This is due to the high crystallinity of Gr, implying that the surface is rich in $sp^2$-bound carbons (basal planes), while the AC has a surface of predominantly edge planes, with dangling bonds terminated by oxygen groups. A low-intensity peak at 292 eV originates from the $\pi$-$\pi^*$ transitions in graphitic carbon[31]. The F 1s spectrum looks identical to AC-pristine, exhibiting a strong peak at 688 eV from PVDF.

In the C 1s spectrum of Gr-12ppm, the atomic percentage of C-$C_{sp^2}$ declines significantly and the $\pi$-$\pi^*$ peak, characteristic of graphitic carbon, vanishes. At the same time, the signals for oxygen-bound carbon (C-O/C=O/O-C=O) gain in atomic percentage. These indicators are characteristic of the SEI formed on Gr, which is created by the reductive decomposition of the carbonate-based electrolyte[55]. Likely, the SEI contains $Li_2CO_3$, which possesses a similar binding energy as the $CF_2$ bond in PVDF[56] and therefore overlays the latter signal. Notably, C-$C_{sp^2}$, located at 284.4 eV in Gr-pristine, shifts to lower binding energies. This is attributed to the formation of $Li_xC_6$, whose peak position can vary from 282 to 284 eV depending on the degree of lithiation of Gr[57]. However, this observation is somewhat unexpected, as the Gr electrodes were electrochemically delithiated prior to measurement (see Fig. 1, step 7). The residual $Li_xC_6$ signal is likely due to trapped $Li_xC_6$ near the surface of the Gr or the carbon black, and it is worth noting that the $Li_xC_6$ peak was also present in delithiated Gr samples in the work of Högström et al.[57]. The peak corresponding to residual $Li_xC_6$ loses intensity in Gr-950ppm and disappears almost entirely in Gr-2300ppm. This could indicate that the SEI thickens as the water concentration increases, consistent with observations from SEM.

In the F 1s spectrum of Gr-12ppm, Gr-950 ppm and Gr-2300 ppm the atomic percentage of the PVDF component is significantly lower compared to Gr-pristine, which is expected as the formation of an SEI that covers the PVDF binder. However, the $CF_2$ component remains clearly visible in the C 1s spectra. This might be explained by the overlap of $Li_2CO_3$ species with the PVDF signal[57,56]. Thus, the seemingly dominant peak at 290.5 eV peak may in fact result from a reduced contribution of $CF_2$ from the PVDF and an increased deposition of $Li_2CO_3$ from solvent decomposition. Additionally, a substantial amount of LiF (at 685.4 eV) is detected in all cycled Gr electrodes, known as a product of the reductive decomposition of $FSI^-$ below 1.9 V vs. Li/$Li^+$[32]. Strikingly, the atomic percentage of LiF is particularly high for Gr-2300ppm, which is in line with the dominant presence of F-rich deposits observed in EDX of Gr-2300ppm (see Fig. S5B). The enhanced formation of LiF in Gr-2300ppm suggests that water contamination participates in or promotes the reductive decomposition of LiFSI.

The fitting and interpretation of S 2p and N 1s spectra was conducted according to the work of Philippe et al.[32]. In the S 2p spectra of all cycled Gr electrodes, the pristine LiFSI salt can be identified ($2p_{3/2}$ = 170.3 eV) alongside its reduction products which result from salt decomposition upon SEI formation[32]. The absolute atomic percentages of LiFSI-red shows a continuous increase from Gr-12ppm over Gr-950ppm to Gr-2300 ppm, suggesting that higher water amounts results in more of the reduction product. However, the atomic percentage of pristine salt increases as well, indicating that some LiFSI is trapped in the SEI or remaining on the surface despite washing with DMC. A third component is also observed ($2p_{3/2}$ = 167.0 eV), which is known as a beam-induced degradation product of LiFSI[32] and is therefore not further considered.

The N 1s spectrum reveals two distinct components, namely the pristine LiFSI salt at 400.5 eV and a reduction product of LiFSI at 399.0 eV[32]. In line with the trend observed in S 2p, the reduction product gains in atomic percentage from Gr-12ppm over Gr-950ppm to Gr-2300 ppm.

In the Li 1s spectrum (see Fig. S8), LiF is detected at 56 eV in all samples, consistent with the pronounced LiF peak in F 1s spectra. There is no peak corresponding to residual $Li_xC_6$ (expected at 53.5 eV[57]), likely due to the low amount of residual $Li_xC_6$ and the low photo-ionization cross-section of the Li 1s core. No peak corresponding to LiOH is observed in Li 1s (expected at 54.5 eV[58]). Neither does the O 1s spectrum (see Fig. S8) show distinct peaks corresponding to LiOH or $Li_2O$ (expected at 531 eV and 528 eV, respectively[58]). The absence of the latter species is interesting as they are expected reaction products of the hydrogen evolution reaction between water and the lithiated Gr electrode. The occurrence of such a reaction would be in line with the rise of the Gr open-circuit potential in the water-containing electrolytes (see Fig 4), which is suspected to stem from irreversible reactions of lithiated Gr and $H_2O$. The lack of LiOH on the surface might indicate that the LiOH is buried deep within the SEI or has detached from the electrode and initiated further reactions, such as for instance the hydrolysis of DMC or EC.

In summary, the post-mortem analysis on Gr showed a moderate increase of SEI thickness for Gr-950 ppm and pronounced gain in SEI thickness in Gr-2300ppm. This suggests that higher amounts of trace water and/or its electrolysis products $H^+/OH^-$ facilitate carbonate decomposition reactions, likely through ring-opening hydrolysis of EC. Furthermore, the prevalence of LiF-based compounds, specifically for the case Gr-2300 ppm, implies that higher water contents promote the electrochemically driven LiFSI hydrolysis. The S 2p and N 1s spectra confirm that LiFSI reduction products are formed in larger quantity as the water amount is increased.

# 5 Electrochemical stability of AC in half-cell setup

Lastly, the effect of water contamination in LiFSI-based electrolyte on the long-term electrochemical stability of AC is closely examined. For this purpose, the AC electrode was subjected to voltage hold (VH) tests. In these tests, AC electrodes were paired with a LFP counter electrode, which was partially delithiated (x = 0.7) to serve as a reference electrode operating at 3.45 V vs. $Li/Li^+$ (see Fig. 2 and Experimental Section). Figure 12 displays the capacity retention of AC when subjected to 0.5 V vs. LFP in LiFSI-12ppm, LiFSI-950ppm, LiFSI-2300ppm, and LiFSI-6000ppm. Also, the capacity retention upon voltage hold at 0.7 V vs. LFP in LiFSI-12ppm is shown to evaluate the positive electrochemical stability limit of the dry LiFSI-based electrolyte on AC.

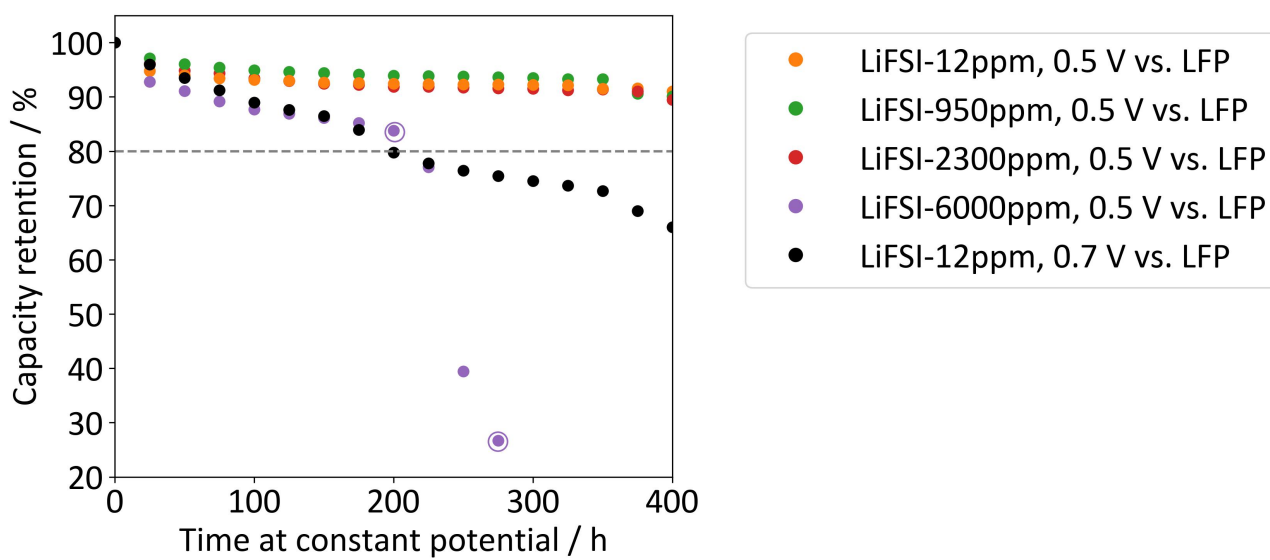

**Figure 12:** Capacity retention of AC electrodes during voltage hold tests in AC/LFP cells.

The AC electrode subjected to 0.5 V vs. LFP (= 3.95 V vs. $Li/Li^+$) in LiFSI-12ppm (orange) shows excellent stability, with over 90% of capacity retention, indicating that this potential is within the electrode's electrochemical stability window. This result aligns with the observations from the LIC full cell in LiFSI-12ppm, where the AC electrode reaches a positive vertex potential of 3.9 V vs. $Li/Li^+$ and exhibits minimal aging. However, at 0.7 V vs. LFP (black), a significant capacity fade to 63% was observed, demonstrating that the AC exceeds its anodic stability limit. This is consistent with previous findings[4] and implies that AC/Gr LICs operating in LiFSI-based carbonate electrolytes should operate at a maximum cell voltage of ∼3.85 V (assuming Gr operates at ∼100 mV vs. $Li/Li^+$) to avoid degradation and capacity fade of the AC positive electrode.

Based on the stable performance of the LiFSI-12ppm electrolyte at 0.5 V vs. LFP, this potential was selected to compare LiFSI-based electrolytes with added water. The LiFSI-950ppm and LiFSI-2300ppm do not lead to significant capacity fade compared to the dry electrolyte. This finding suggests that trace water at these concentrations does not adversely affect the electrochemical stability of the AC electrode. In LiFSI-6000ppm, the AC electrode exhibited continuous capacity fade, with a marked decline after 200 hours of holding time, indicating that a water content as high as 6000 ppm promotes parasitic reactions and severely compromises electrochemical stability.

Fig. 13 shows post-mortem SEM images and F 1s XPS spectra of AC electrodes held at 0.5 V vs. LFP. The nomenclature (e.g., AC-VH-950ppm) specifies the water content in the electrolyte, with voltage hold times of 400 hours. Due to the pronounced capacity fade in LiFSI-6000ppm, the AC electrodes exposed to this electrolyte were retrieved after 200 hours and 275 hours of voltage hold duration (see encircled data points in Fig. 12).

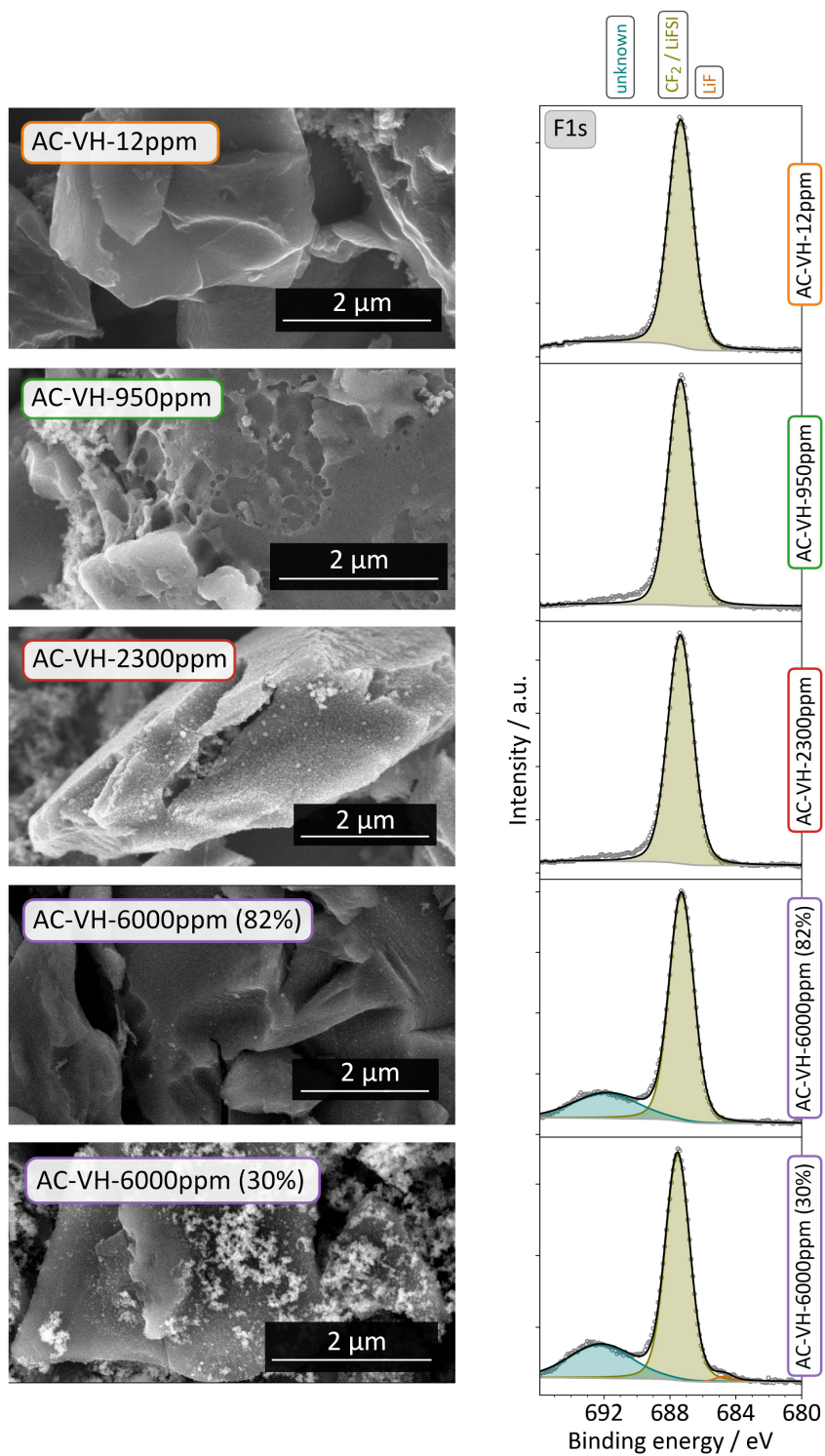

**Figure 13:** Post-mortem SEM and F 1s spectra of AC-VH-12ppm, AC-VH-950ppm, AC-VH-2300ppm, AC-VH-6000ppm (82%), AC-VH-6000ppm (30%). The values in brackets refer to the capacity retention of AC, i.e., 82% after 200 hours and 30% after 275 hours.

The SEM image of AC-VH-12ppm reveals a surface morphology nearly identical to that of pristine AC, with no detectable deposits. This is in line with the high capacity retention (92%) and shows that the AC electrode remains stable in the presence of low water content. No LiF is detected, similar to AC-12ppm in LIC full cells which revealed only a small amount of LiF (see Fig. 8).

The surface of AC-VH-950ppm appears clean and free of significant deposits, a result that together with the high capacity retention (91%) suggests good compatibility of AC with 950 ppm water upon long-term exposure to 0.5 V vs. LFP (equivalent to 3.95 V vs. $Li/Li^+$). In fact, the electrode surface lacks LiF, a species which has been observed on AC-950ppm in LIC full cells (see Fig. 8). This is somewhat unexpected, as the positive vertex potentials are similar in galvanostatic cycling of LICs and during voltage hold of AC/LFP cells ($\sim$3.9 V vs. $Li/Li^+$ and 3.95 V vs. $Li/Li^+$, respectively). Moreover, voltage hold tests are expected to accelerate the aging of AC compared to galvanostatic cycling[30]. The discrepancy might be attributed to crosstalk effects where $FSI^-$ decomposition is influenced by interactions between trace water and the lithiated Gr negative electrode. A possible mechanism is the formation of hydroxide ions at the Gr surface, which could attack $FSI^-$ ions, leading to $F^-$ abstraction. However, the different cell architectures and the hydrophilic nature of LFP may impact the degradation patterns, possibly reducing the extent of AC aging in AC/LFP voltage hold tests compared to LIC full cells. A direct comparison between the AC/LFP half cells and Gr/AC full cells must therefore be treated with caution.

The surface of AC-VH-2300ppm shows widespread, grainy deposits across the electrode surface. Spot-EDX analysis (see Fig. S9A) identifies these deposits as rich in phosphorous (P), suggesting that water induces significant phosphate dissolution from the LFP counter electrode and re-deposition on AC. This interference from phosphate dissolution complicates the interpretation of the AC electrode's interfacial aging processes in the presence of 2300 ppm of water. Similar deposits containing P, O, and F are present on AC-VH-6000ppm (82%), with increased coverage seen on AC-VH-6000ppm (30%), indicating that the severity of phosphate deposition rises with higher water content (see Fig. 13 for EDX mapping). This extensive phosphate coverage likely contributes to the marked capacity fade of AC in LiFSI-6000ppm by blocking the AC's pore network. Despite the dominant side process of phosphate deposition, it is noteworthy that LiF is absent in AC-VH-2300ppm. Only the very degraded AC-VH-6000ppm electrode shows a small amount of LiF. This further suggests that the crosstalk effect with the lithiated Gr negative electrode is primarily responsible for decomposition of $FSI^-$ and the subsequent formation of LiF on the AC surface.

The strong interference of phosphate deposition on AC's surface necessitates complementary approaches to explain the aging processes in the cell exposed to LiFSI-6000ppm. To investigate this, the electrolytes from cells aged in LiFSI-6000ppm for 200 h (82% capacity retention) and 275 h (30% capacity retention) were recovered and analyzed using $^1$H NMR (Fig. 14B and C). For comparison, the $^1$H NMR spectrum of an uncycled LiFSI-12ppm electrolyte is shown in Fig. 14A. Signals at 3.18 and 4.08 ppm are attributed to methanol[59], a known hydrolysis product of DMC. The occurrence of these signals in Fig. 14B, and the intensity increase of these signals in Fig. 14C suggests that the capacity fade of AC is associated with an increased propensity for DMC hydrolysis. Additionally, a low-intensity signal at 8.13 ppm is detected in both Fig. 14B and C, indicative of formic acid[60], which forms through oxidation of methanol generated by hydrolysis of DMC. A signal at 9.65 ppm, attributed to formaldehyde[60], another oxidation product of methanol, is present only in Fig. 14C, linking the accelerated capacity fade with enhanced formaldehyde formation. Finally, an increase of various unidentified signals between 6.95 and 7.46 ppm is observed in Fig. 14C, speculatively originating from aromatic organic compounds. This demonstrates that the capacity fade in LiFSI-6000ppm is accompanied by the formation of several additional solvent decomposition products.

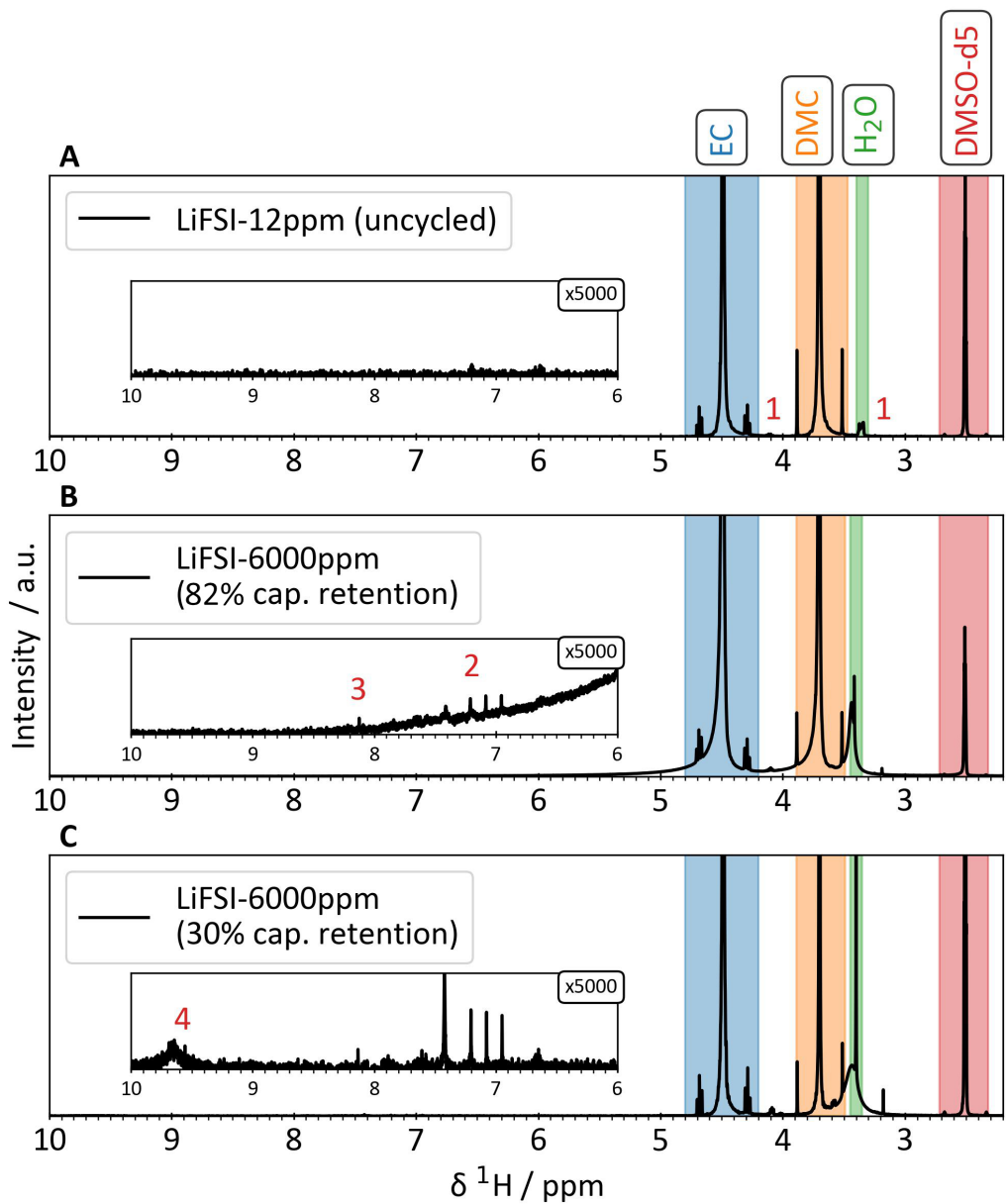

**Figure 14:** $^1$H NMR spectra of **A)** uncycled LiFSI-12ppm electrolyte, **B)** retrieved electrolyte after voltage hold of AC/LFP half cells at 0.5 V vs. LFP in LiFSI-6000ppm with 82% capacity retention, **C)** retrieved electrolyte after voltage hold of AC/LFP half cells at 0.5 V vs. LFP in LiFSI-6000ppm with 30% capacity retention. Signal assignments: 3.18 & 4.08 ppm: methanol ("1"); 6.95, 7.05, 7.3 & 7.45 ppm: unknown ("2"); 8.13 ppm: formic acid ("3"); 9.65 ppm: formaldehyde ("4").

# 6 Conclusions

This study systematically investigated the cycling stability and aging of an AC/Gr-based LIC when a deliberate amount of trace water was added to an EC/DMC + 1.0 M LiFSI electrolyte. Cycling the LIC for 2000 cycles in the dry electrolyte (<12 ppm $H_2O$) did not induce noteworthy signs of degradation on either AC or Gr. Increasing the trace water concentration triggered the following phenomena on the lithiated Gr side, in progressive severity: (1) Irreversible side reactions of water on the Gr surface/bulk, consuming lithium inventory. This led to a positive drift of the Gr electrode's operating potential, destabilizing the performance and causing capacity fade. A water concentration of 6000 ppm cancels out the beneficial effect of pre-lithiation. (2) Thickening of the SEI on Gr, caused by the aggravation of carbonate hydrolysis and polymerization, likely catalyzed through the water electrolysis products $H^+$/$OH^-$. (3) Increased $FSI^-$ reductive decomposition, yielding widespread deposits of $Li_xO_yF_z$, particularly for electrolytes containing 2300 ppm and 6000 ppm $H_2O$.

On the AC side in LIC full cells, rising water concentrations increase the abundance of spherical $Li_xO_yF_z$ deposits, suggesting that water is involved in and aggravates the oxidative decomposition of LiFSI. However, up to a water concentration of 2300 ppm, the oxidative decomposition of the solvent and subsequent polymerization on AC occurs to a lower extent compared to Gr, producing a surface film too thin to be detected by SEM and only suggested by XPS. The aging studies of AC/LFP cells involving voltage hold tests at 0.5 V vs. LFP ( = 3.95 V vs. $Li/Li^+$) revealed that the capacity fade of the AC electrode was minor up to a water concentration of 2300 ppm, and the lack of $Li_xO_yF_z$ deposits suggests that the decomposition of $FSI^-$ is accelerated by crosstalk effects of the water with the lithiated Gr electrode. However, the dissolution and re-deposition of the LFP counter electrode complicates the interpretation of the aging processes at the AC/electrolyte interphase.

Despite the obvious detrimental effects of trace water on LIC performance, it is worth highlighting that the LIC cycled in electrolyte with 950 ppm $H_2O$ achieved the same cycling performance as the dry electrolyte (96% after 2000 cycles), while sustaining only subtle morphological and chemical changes on AC and Gr. Also, the AC-LFP cells delivered excellent stability of AC during long-term aging at 950 ppm water. It can be argued that the water "tolerance" of such an LIC reaches up to 950 ppm, meaning satisfactory device performance can be maintained despite such high water levels. This makes LiFSI very promising compared to conventional $LiPF_6$ based electrolytes: $LiPF_6$-based electrolytes are highly unstable in the presence of water. In contrast, LiFSI-based electrolytes not only offer excellent shelf life due to their inertness to water at room temperature but also demonstrate impressive cycling performance even at water levels up to 950 ppm. This makes LiFSI a major advancement toward reducing costs of electrolyte drying in LIC production. Further studies should explore the additive use of water scavenger molecules such as LiTDI, which may increase the water tolerance even further by preventing the contact of water with the electrochemical interfaces.

# Acknowledgements

The authors gratefully acknowledge funding from the Research Council of Norway (NFR) under ENERGIX programme no. 306400 as part of the "Norwegian Giga Battery Factories" (NorGiBatF) project.

# Author contributions

**Philipp Schweigart**: Data curation: Lead; Formal analysis: Lead; Investigation: Lead; Writing – original draft: Lead; Writing – review & editing: Lead)

**Dr. Johan Hamonnet**: Investigation: Supporting; Writing – original draft: Equal; Writing – review & editing: Equal

**Dr. Obinna Egwu Eleri**: Investigation: Supporting; Writing – original draft: Equal; Writing – review & editing: Equal.

**Dr. Samson Yuxiu Lai**: Investigation: Supporting; Writing – original draft: Supporting; Writing – review & editing: Equal

**Laura King**: Investigation: Supporting (XPS acquisition)

**Prof. Ann Mari Svensson**: Conceptualization: Lead; Funding acquisition: Lead; Investigation: Lead; Supervision: Lead; Writing – original draft: Lead; Writing – review & editing: Lead)

# Conflict of Interest

The authors declare no conflict of interest.

# Keywords

Li-ion capacitor · Activated carbon · Graphite · Aging · Water contamination · Moisture · LiFSI

# The Effect of Water Contamination on the Aging of a Dual-Carbon Lithium-Ion Capacitor Employing LiFSI-based Electrolyte

## Supporting Information

Philipp Schweigart[1], Johan Hamonnet[1], Obinna Egwu Eleri[2], Laura King[3], Samson Yuxiu Lai[4], and Ann Mari Svensson[1]

[1]Department of Materials Science and Engineering, Norwegian University of Science and Technology, Trondheim, Norway
[2]Beyonder AS, Sandnes, Norway
[3]Department of Chemistry - Ångström Laboratory, Uppsala University, Uppsala, Sweden
[4]Battery Technology Department, Institute for Energy Technology, Kjeller, Norway

**Corresponding authors**
PS: philipp.schweigart@ntnu.no
AS: annmari.svensson@ntnu.no

# 1 Carbon-primed current collector for AC electrodes

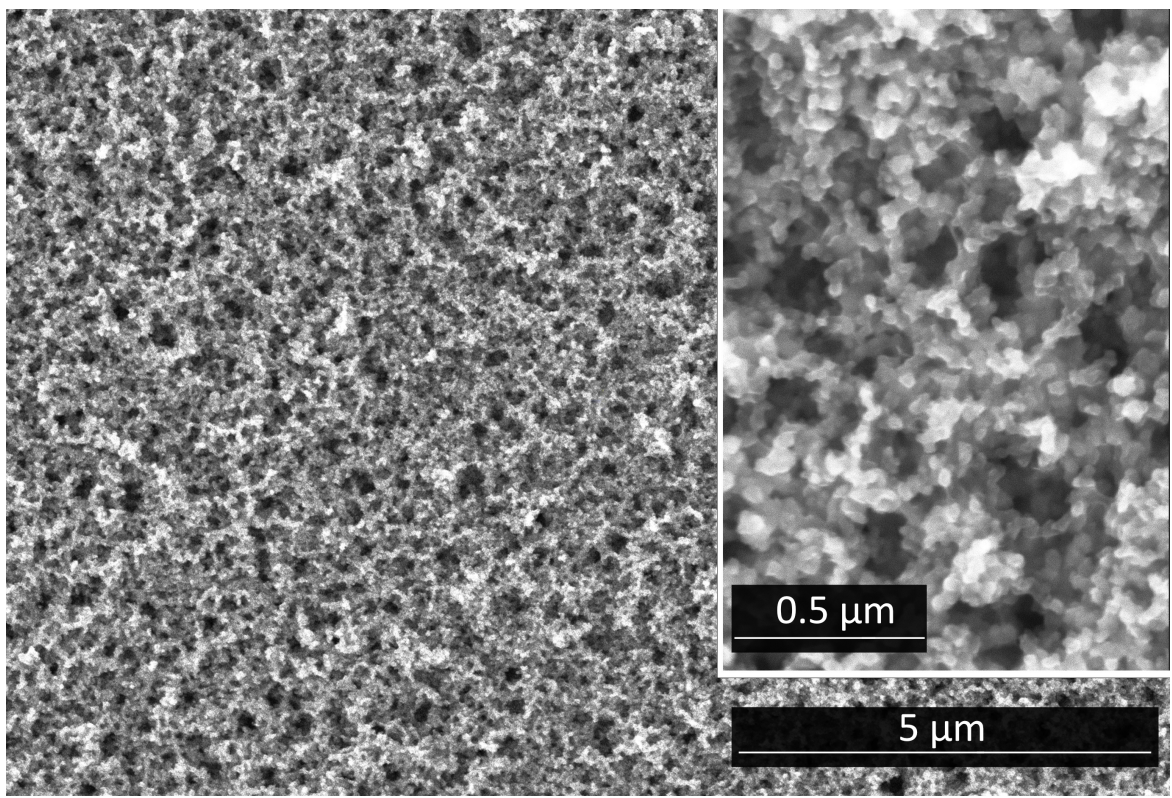

**Figure S1:** Electron micrograph of Ensafe20 double-sided carbon-primed aluminum foil used as current collector for the AC positive electrodes.

# 2 Recovery of electrolyte for post-mortem NMR

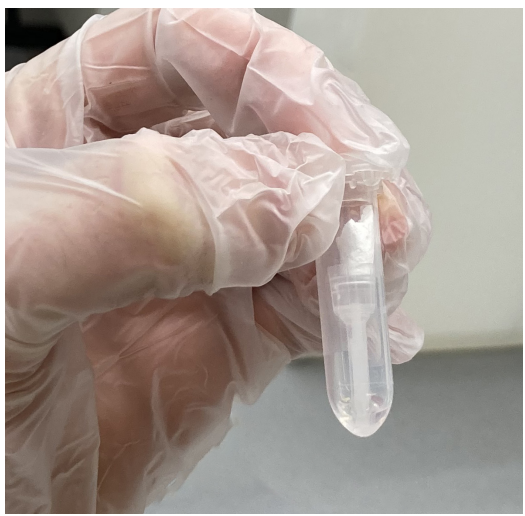

**Figure S2:** Recovery of electrolytes from aged cells. The glass fiber separator extracted from the cell is placed in a safe-lock tube equipped with a pipette tip. After centrifugation (6000 rpm, 20 min), the electrolyte accumulates at the bottom of the safe-lock tube.

## 3 Absolute capacities of LICs

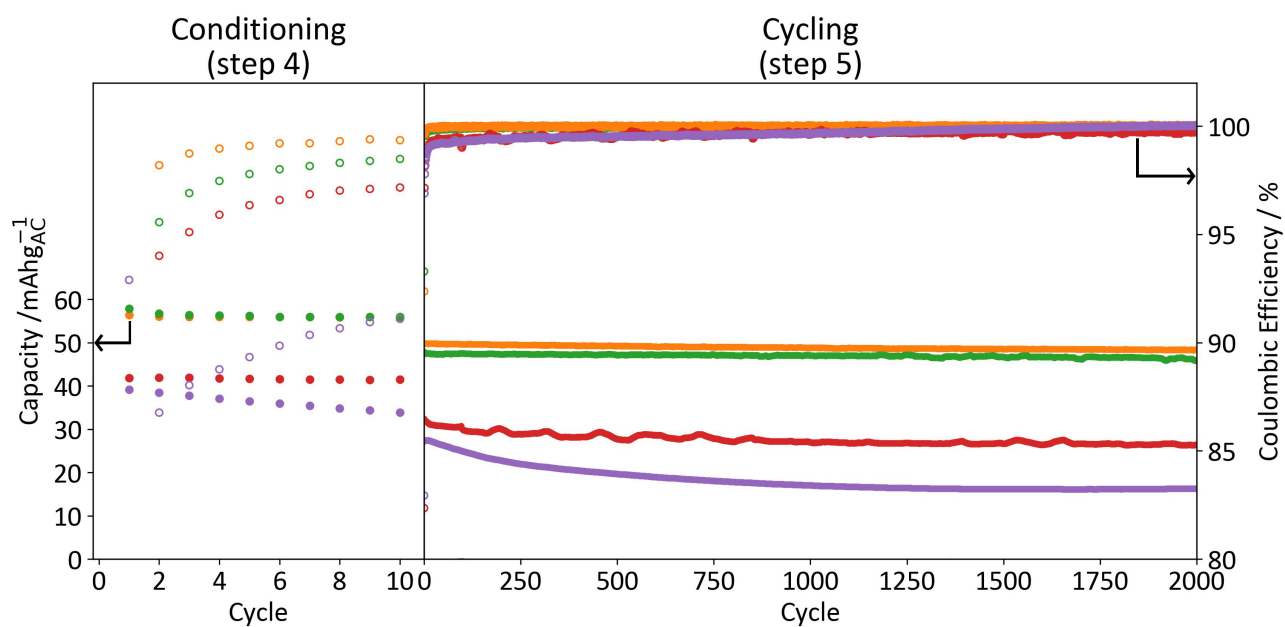

**Figure S3:** Absolute capacities of LICs normalized to AC active material mass (filled circles) and coulombic efficiencies (open circles) during conditioning and cycling steps.

# 4 EDX elemental mappings

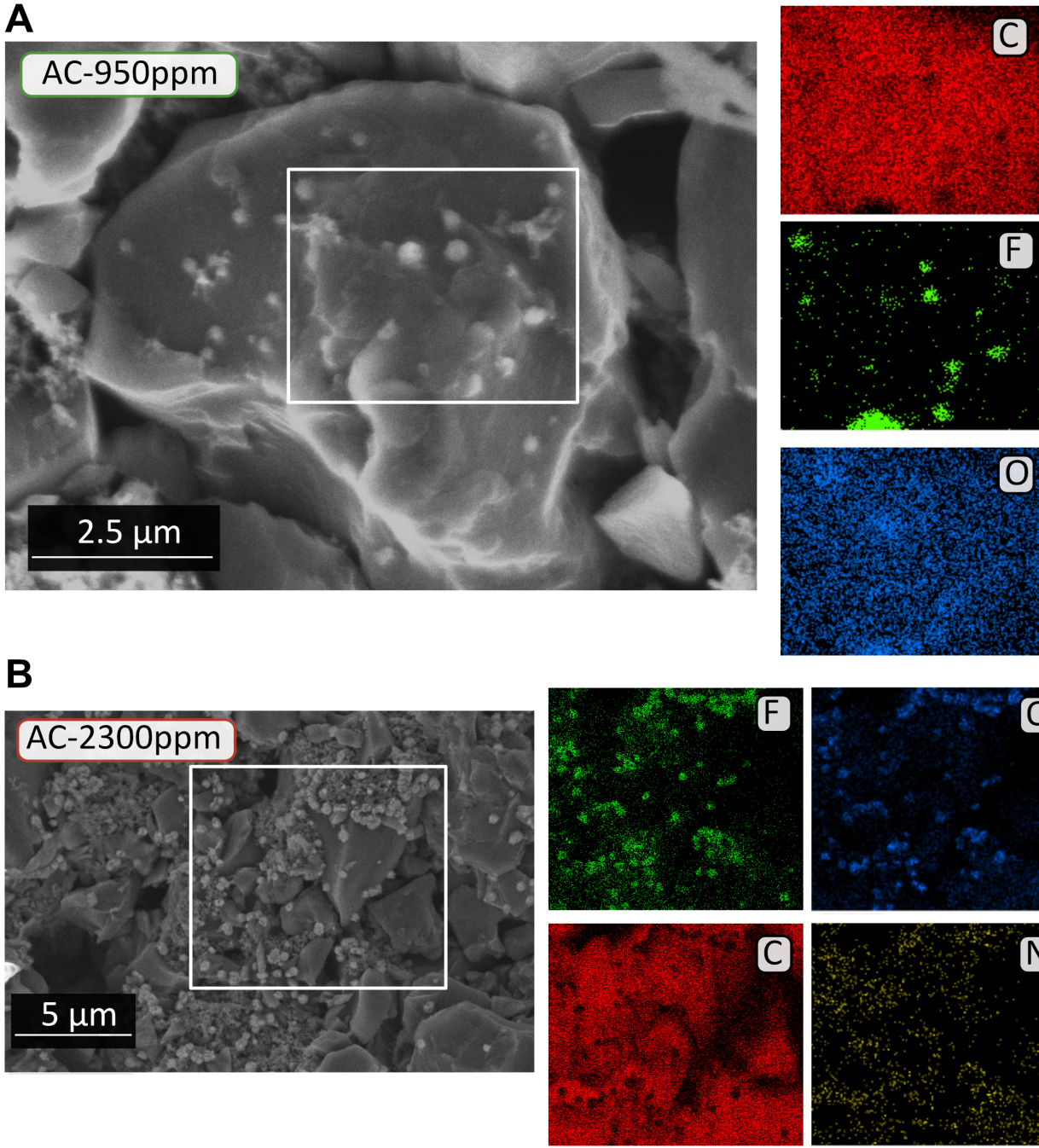

**Figure S4:** EDX elemental mappings of **A)** AC-950ppm **B)** AC-2300ppm.

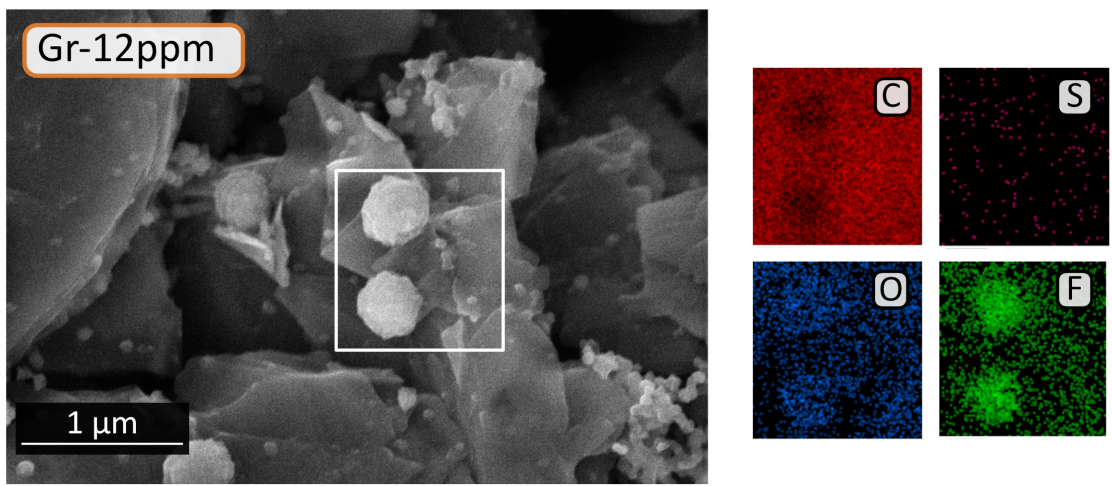

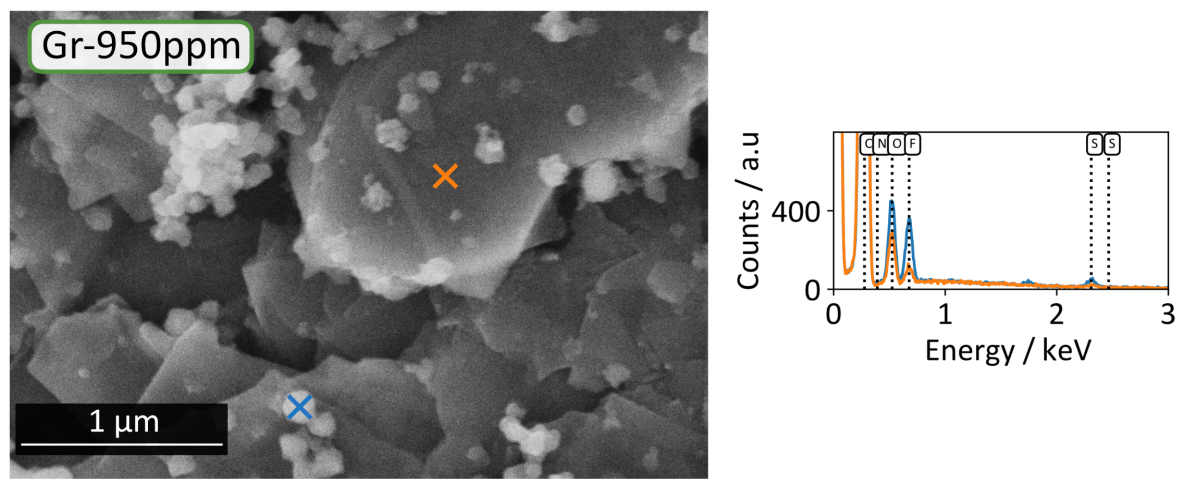

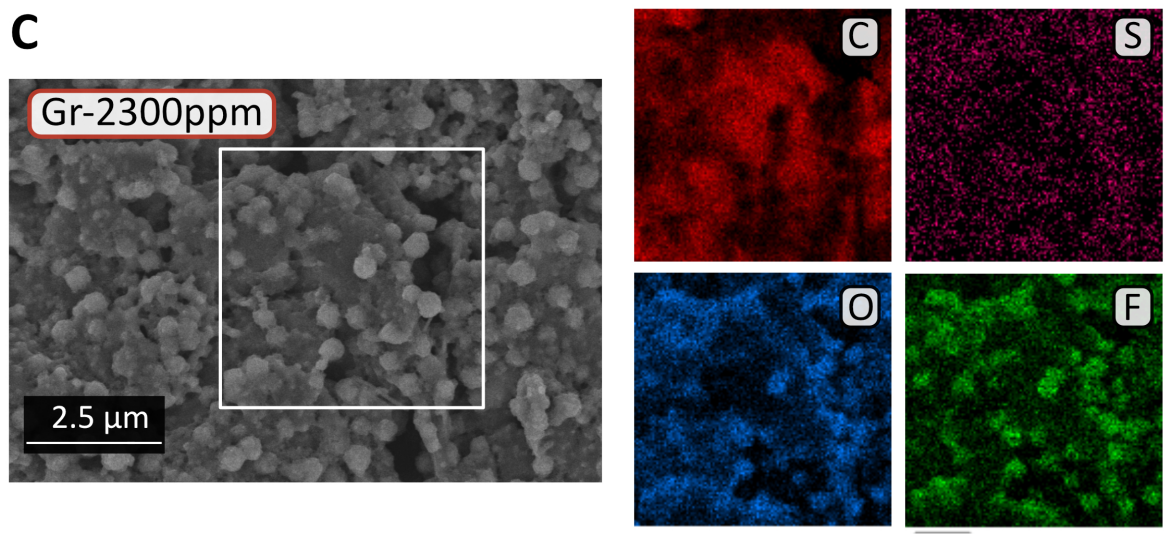

**Figure S5: A)** EDX elemental mappings of Gr-12ppm **B)** spot-EDX of Gr-950ppm **C)** EDX elemental mapping of Gr-2300ppm.

# 5 XPS survey spectra

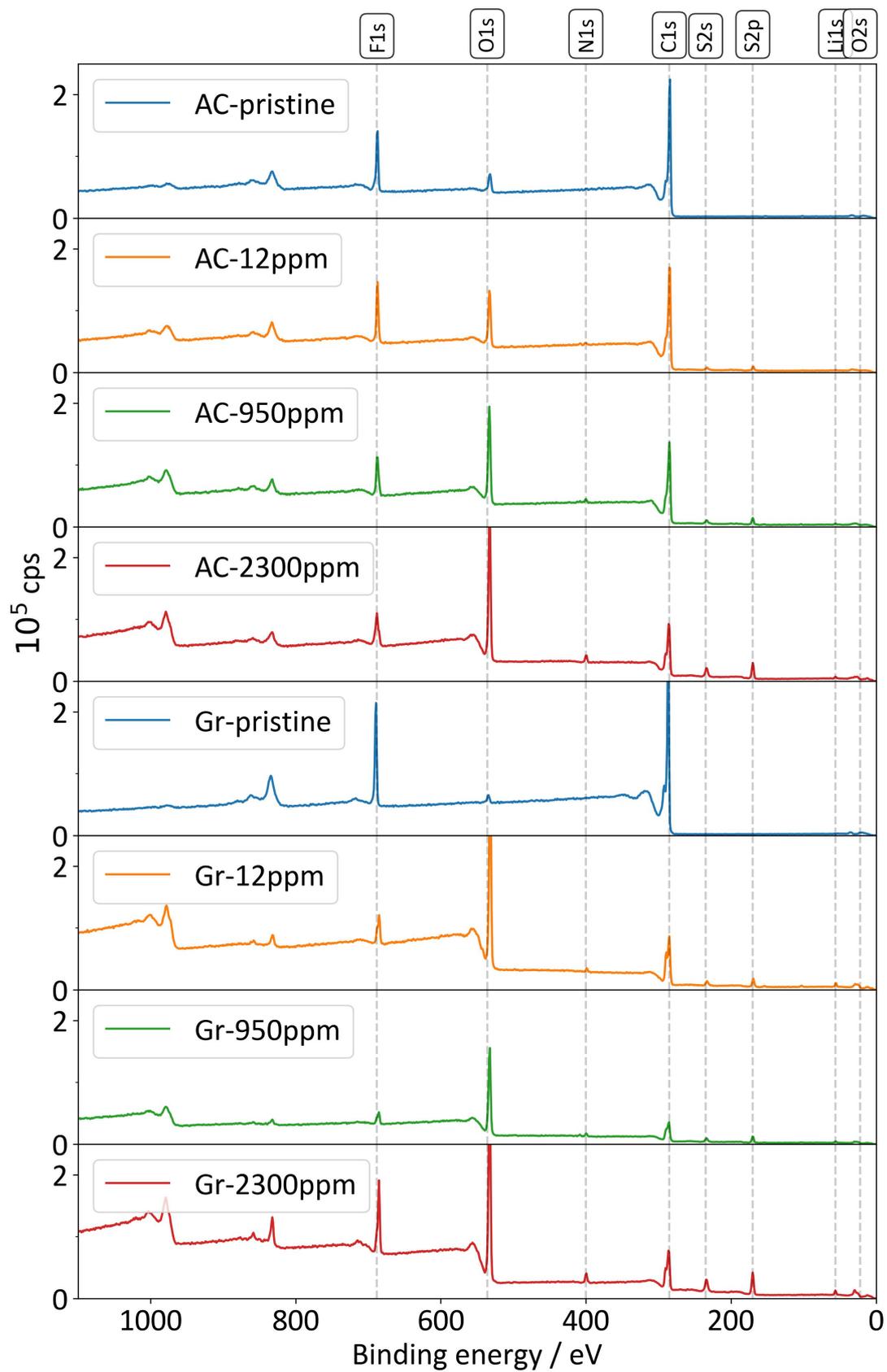

**Figure S6:** XPS survey spectra.

# 6 Additonal XPS spectra

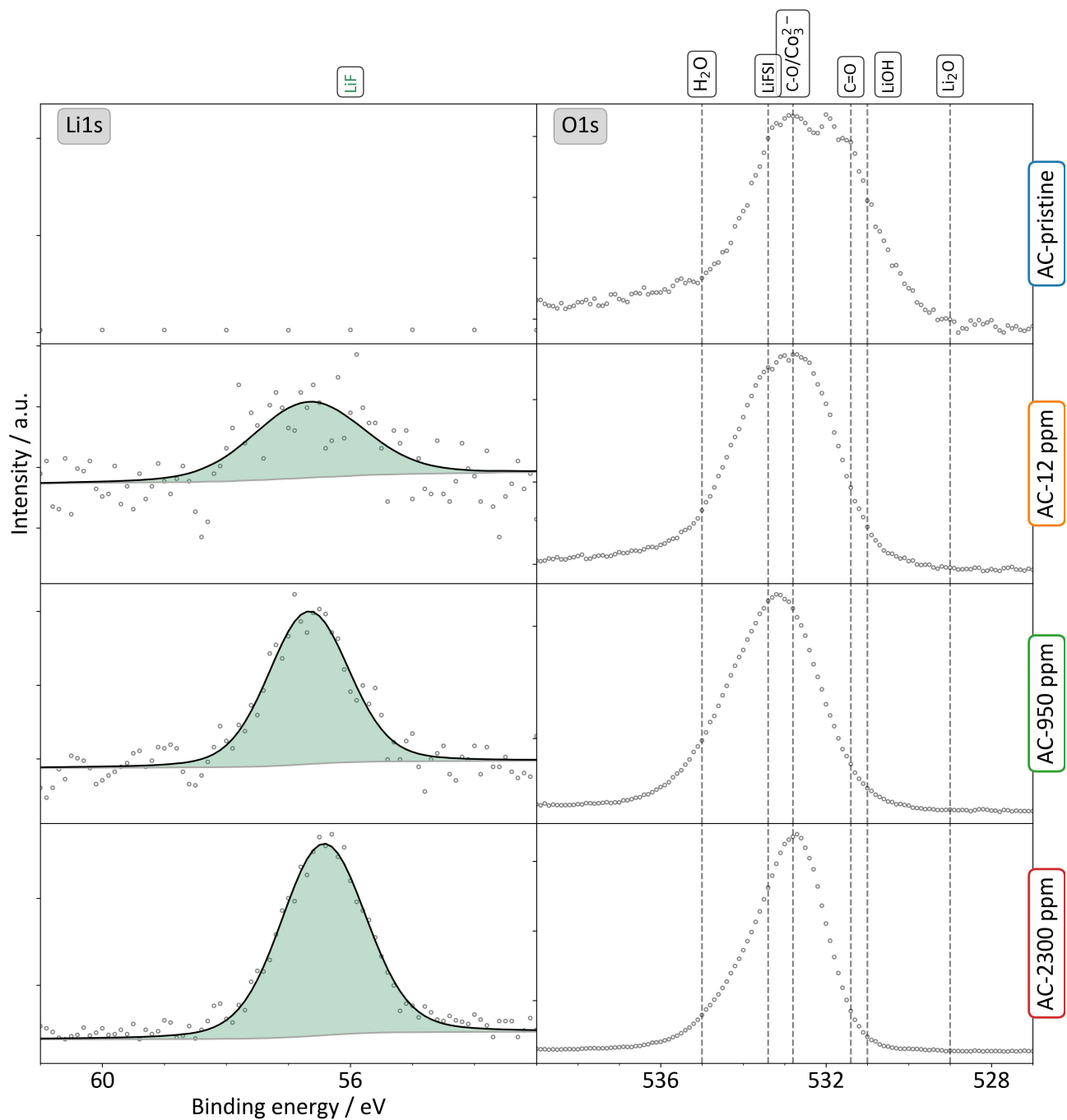

**Figure S7:** Li1s and O1s spectra of AC-pristine, AC-12ppm, AC-950ppm and AC-2300ppm.

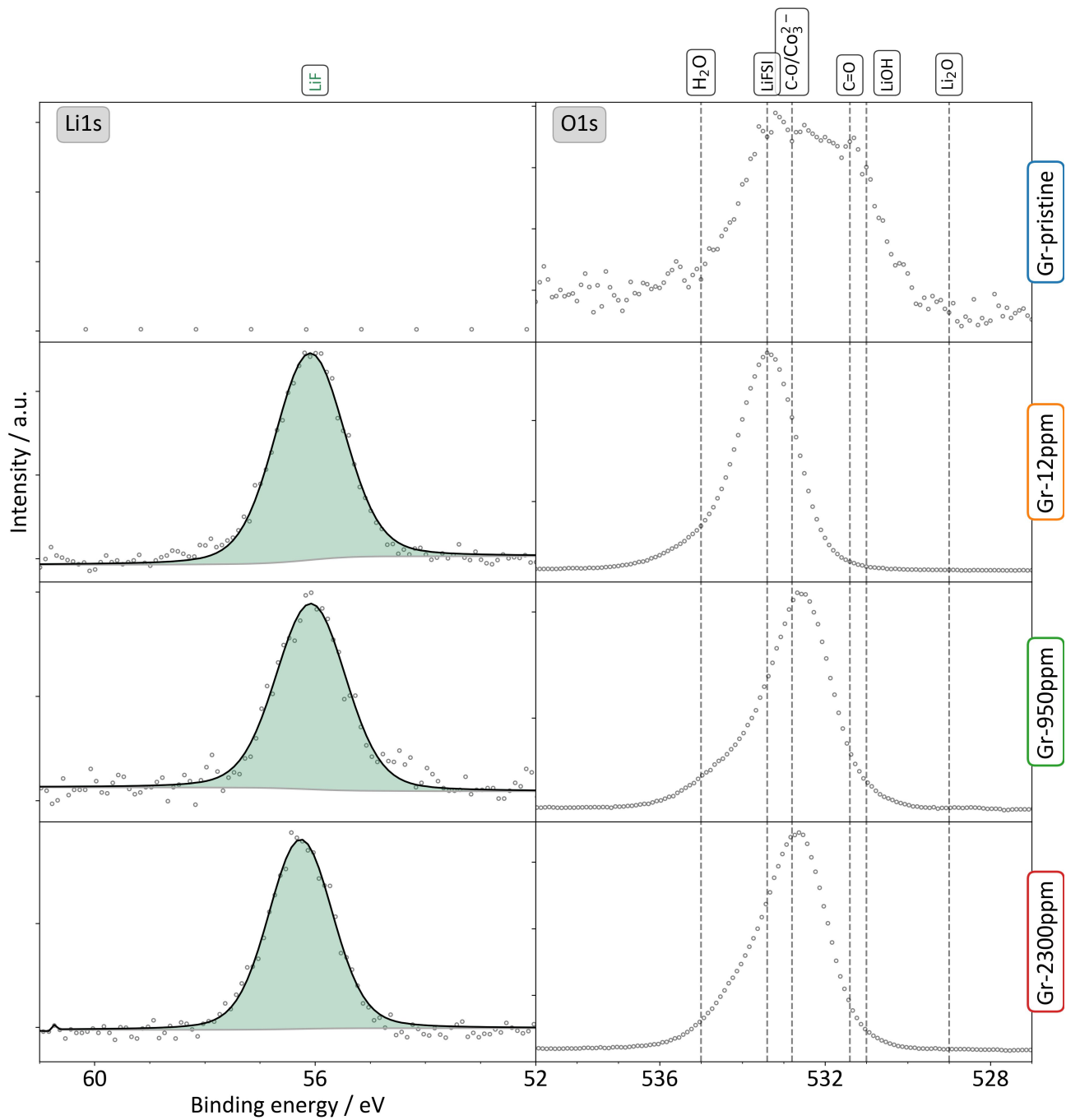

**Figure S8:** Li1s and O1s spectra of Gr-pristine, Gr-12ppm, Gr-950ppm and Gr-2300ppm.

## 7 EDX mappings of AC electrodes from AC/LFP cells

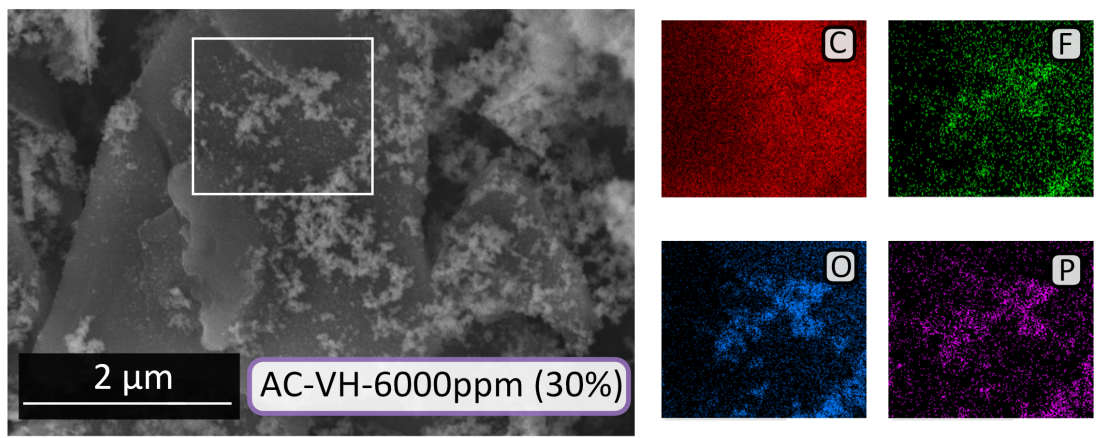

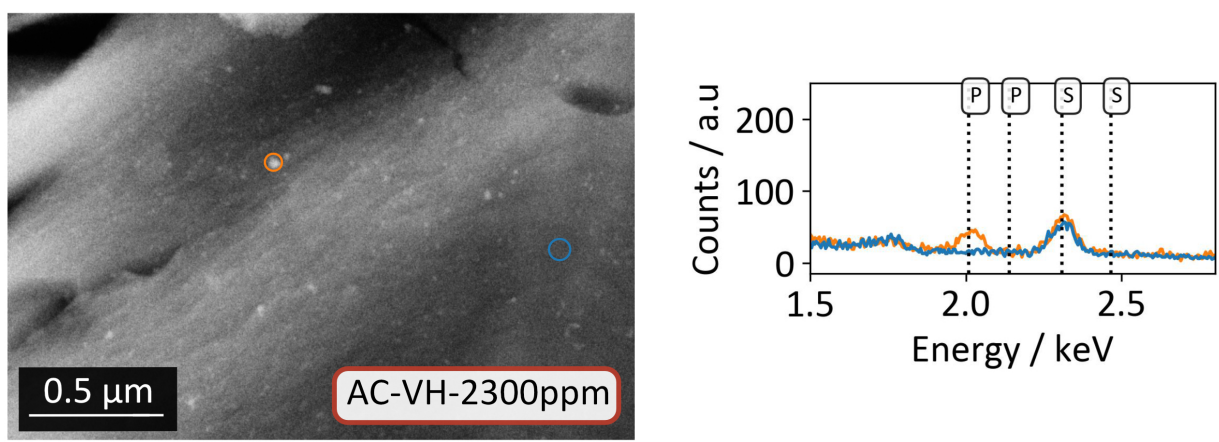

**Figure S9:** Post-mortem EDX mappings of **A)** AC-VH-6000ppm (30%), **B)** AC-VH-2300ppm.